\documentclass[12pt,english]{article}

\usepackage[utf8]{inputenc}

\usepackage[left=1in,right=1in,top=1in,bottom=1in]{geometry}

\usepackage{graphicx}
\usepackage{subcaption}
\usepackage{braket}
\usepackage{amsmath}
\usepackage{amssymb}

\usepackage{physics}
\usepackage{leftidx}
\usepackage{feynmf}
\usepackage{wrapfig} 
\usepackage{authblk}

\usepackage{prettyref}

\usepackage{esint}

\usepackage{float}

\usepackage{dsfont}

\usepackage{xcolor}
\usepackage{color}
\definecolor{darkgreen}{rgb}{0,0.5,0}
\definecolor{darkblue}{rgb}{0,0,0.6}
\definecolor{purple}{rgb}{0.4,.2,0.7}
\usepackage[colorlinks=true,citecolor=darkgreen,linkcolor=purple,urlcolor=purple]{hyperref}

\usepackage{enumitem}
\usepackage{ulem} 
\usepackage{tensor} 
\usepackage{slashed}
\usepackage{mathtools}
\usepackage{empheq}

\def\nn{\nonumber}

\def\qs{ \bar{q}\left (\frac{Q}{2M}-aB\right )}

\numberwithin{equation}{section}
\numberwithin{table}{section}

\begin{document}

\title{Penrose process for a charged black hole in a uniform magnetic field}  


\author{\small Kshitij Gupta}
\author{Y.T.\ Albert Law}
\affil{Department of Physics, Columbia University, New York, NY 10027}

\author[2]{Janna Levin}
\affil{Department of Physics and Astronomy, Barnard College of Columbia University, New York, NY 10027}
\date{}

\maketitle




\begin{center}
{\footnotesize  Email: kg2789@columbia.edu, yal2109@columbia.edu, janna@astro.columbia.edu}
\end{center}

\vskip2cm

\vskip0mm

\thispagestyle{empty}
\begin{abstract}
    Spinning black holes create electromagnetic storms when immersed in ambient magnetic fields, illuminating the otherwise epically dark terrain. In an electromagnetic extension of the Penrose process, tremendous energy can be extracted, boosting the energy of radiating particles far more efficiently than the mechanical Penrose process. We locate the regions from which energy can be mined and demonstrate explicitly that they are no longer restricted to the ergosphere. We also show that there can be toroidal regions that trap negative energy particles in orbit around the black hole. We find that the effective charge coupling between the black hole and the super-radiant particles decreases as energy is extracted, much like the spin of a black hole decreases in the mechanical analogue. While the effective coupling decreases, the actual charge of the black hole increases in magnitude reaching the energetically-favored Wald value, at which point energy extraction is impeded. We demonstrate the array of orbits for products from the electromagnetic Penrose process.
\end{abstract}


\newpage
\thispagestyle{empty}
\tableofcontents


\pagenumbering{arabic} 

\section{INTRODUCTION}

The 21st century has been remarkable for black hole discoveries, from LIGO's first recording of the collision of two black holes \cite{LIGO} to the EHT image of the shadow cast by a black hole event horizon \cite{EHT}. Loud black hole mergers seem to occur in total darkness while the Event Horizon Telescope has the potential to capture silent movies. In the composite story that emerges, black holes have asserted themselves as plentiful -- there are many more across a vast range of masses than previously predicted -- and as influential -- the supermassive black holes are the sculptors of their galaxies. 

Although intrinsically the darkest phenomena conceivable in the universe, black holes are famously also the single most powerful luminous engines conceivable, creating electromagnetic storms in the surrounding environment and driving jets powerful enough to blow holes in neighboring galaxies.

Black hole engines occur when these otherwise empty locales are immersed in external magnetic fields, which are transported by neutron stars, for instance, or threaded through orbiting debris. Whenever a spinning black hole churns up an ambient magnetic field, there is an opportunity for ultra-powerful boosts in energy through an electromagnetic Penrose process, sometimes called the magnetic Penrose process \cite{1985ApJ...290...12W,1986ApJ...307...38P,1985JApA....6...85B}. A classic review of the subject is \cite{WAGH1989137}. An updated exposition covering the subsequent decades of developments can be found in \cite{Tursunov:2019oiq}.

In the purely geometric mechanical Penrose process \cite{Penrose:1969, Penrose}, absent any electromagnetic fields, an outgoing particle gets a boost in energy by cleverly exploiting the relativity of space and time. The energy comes at the expense of the spin of the black hole and occurs solely within the ergosphere with a maximum efficiency of roughly $20\%$.

By contrast, electromagnetic super-radiance leverages the tremendous store of energy in the electromagnetic fields and can lead to ultrahigh efficiencies \cite{Dadhich:2018gmh}, far greater than those of the mechanical process.  Consequently, the electromagnetic Penrose process is a compelling explanation for high-energy astrophysical phenomena such as ultrahigh-energy cosmic rays (UHECRs), particles \cite{Kop_ek_2018,Ruffini:2018aiq,Tursunov:2020juz} observed with energy of about $10^{18}$eV \cite{PierreAuger:2018qvk,PierreAuger:2017pzq}, or relativistic jets \cite{Stuchlik:2015nlt,Istomin:2020oja}. To be explicit about terminology, we are envisioning a process typified by (but not restricted to) the decay of a particle near a black hole that results in a negative-energy daughter and a positive-energy daughter that is radiated with more energy than the parent. We will use the terminology that the positive-energy daughter is superradiant. We can also call the process an electromagnetic Penrose process even though the mechanism may be exploiting electromagnetic interactions and may not depend solely on the relativity of spacetime. Perhaps none of the terminologies are ideal, but we rely on them for brevity.

Interestingly, the electromagnetic Penrose process is related to the Wald charge \cite{Wald:1974np}, which is the natural charge favored for a black hole  when spinning in an ambient magnetic field. While there is a presumption that charge swiftly neutralizes in astrophysical settings, Wald \cite{Wald:1974np} proved that for a spinning black hole in a uniform magnetic field, the opposite is true: Black holes are inclined to charge up. The energetically favorable value of the charge of a black hole is given by the Wald value, which in an external field $B$ for a black hole of mass $M$ and spin $a$ is $Q_W=2aMB$ \cite{Wald:1974np}.

We show that the magnitude of the energy boost that can be delivered to an outgoing particle through the electromagnetic Penrose process is set roughly by the combination
\begin{equation}
    \chi_Q=\qs
\end{equation}
where $\bar q$ is the charge per unit mass of a particle around the black hole. As we argue, the combination $\chi_Q$ summarizes the effective charge coupling between black hole and particle. If we restrict $0\le |Q|\le |Q_W|$, then
the energy extraction is largest when the effective charge coupling, $\chi_Q$, is most negative, which is actually for an uncharged black hole ($\chi_Q=-\bar q a B,Q=0$), and decreases as the Penrose process charges the black hole up to the Wald value ($\chi_Q=0,Q=Q_W$). The decrease in the effective coupling $\chi_Q$ through the electromagnetic Penrose process is analogous to the  slow down of the spin of a black hole through the mechanical Penrose process.

We explicitly locate the regions from which energy can be mined and show that the electromagnetic Penrose process is not restricted to the ergosphere, first observed in \cite{PhysRevD.29.2712,PhysRevD.30.1625}. There can even be disconnected, toroidal regions in which energy can be extracted, to our knowledge the first demonstration of its kind. Within these toroidal regions, negative energy particles are forever trapped, unable to fall into the black hole or to escape.

As we discuss, natural values of the effective coupling are enormous, $\chi_Q\sim 10^{10}-10^{21}$, leading to dramatic boosts in power. The implications for black hole batteries \cite{McWilliams:2011zi, DOrazio:2013ngp, DOrazio:2015jcb, Mingarelli_2015} as well as black hole powered jets \cite{10.1093/mnras/179.3.433} may be significant.

Black hole batteries form when a neutron star threads a companion black hole with its substantial dipole field. By whipping around the neutron star magnet in the final stages before swallowing the star whole, the black hole powers a battery that can light up the system for a luminous complement to a gravitational-wave detection \cite{McWilliams:2011zi, DOrazio:2013ngp, DOrazio:2015jcb, Mingarelli_2015}. If the black hole acquires charge through the Wald mechanism, then a black hole pulsar can also form, if briefly and erratically \cite{Levin:2018mzg}. The electromagnetic process we investigate here can lead to ultra-efficient power boosts to both of these compelling signatures.

For supermassive black holes, the efficient boost in power near the event horizon and even along the jets could be observable to the Event Horizon Telescope project given their detailed observations of M87*, the black hole 6.5 million times the mass of the sun in the neighboring M87 galaxy, 55 million light-years away.

Whether a system will avail itself of these substantial boosts in power depends on the detailed collisional and decay processes fluxing around the black hole. While it is beyond the scope of this work to investigate those rates, we look forward to future assessments of the importance of the generalized Penrose process in a realistic numerical modelling of a black hole environment.


\section{THE BLACK HOLE AND THE ELECTROMAGNETIC ENVIRONMENT}

We will take the electromagnetic energy density to be small enough that the Kerr vacuum solution is valid. Although negligible in terms of modifying the metric, the electromagnetic fields have a significant effect on the dynamics of charged particles in the black hole spacetime, in particular for our interests, on the Penrose process. The metric is then
\begin{align}\label{Kerr metric BL}
ds^2=&-\left (1-\frac{2M r}{\Sigma}\right )dt^2+\frac{\Sigma}{\Delta}dr^2+\Sigma d\theta^2 \nn\\
&+\frac{(r^2+a^2)^2-\Delta a^2 \sin^2 \theta}{\Sigma}\sin^2 \theta d\phi^2
-\frac{2a(2M r)\sin^2 \theta}{\Sigma}dt d\phi
\end{align}
with 
\begin{align}
\Sigma &=r^2 +a^2 \cos^2 \theta \ , \quad\quad \Delta =r^2+a^2-2Mr =(r-r_+)(r-r_-) \quad .
\end{align}
Here $r_+$ and $r_-$ are the positions of the outer and inner horizons respectively:
\begin{align}
r_\pm = M \pm \sqrt{M^2 -a^2}
\end{align}
which satisfy elementary relations like
\begin{align}
r_+ +r_- =2M,\quad r_+ r_- =a^2.
\end{align}
Other useful relations include:
\begin{gather}
g_{t\phi}^2-g_{tt}g_{\phi\phi} = \Delta \sin^2\theta,\quad g^{rr}=g_{rr}^{-1},g^{\theta \theta}=g_{\theta \theta}^{-1},\nn \\ g^{tt}=-g_{\phi\phi}/(\Delta \sin^2\theta),\quad g^{t\phi}=g_{t\phi}/(\Delta \sin^2\theta), \quad g^{\phi\phi}=-g_{tt}/(\Delta \sin^2\theta). 
\end{gather}

In a pure Kerr geometry without any external electromagnetic fields, the mechanical Penrose process occurs inside the ergosphere, which is the region $r_e>r>r_+$ bounded by the stationary surface with $g_{tt} =0$, which occurs at
\begin{align}
r_e (\theta) = M + \sqrt{M^2 -a^2\cos^2\theta}
\end{align}
with $2M > r_e \ge r_+$. 

The vector potential of a spinning, charged black hole aligned with an asymptotically uniform magnetic field is \cite{Wald:1974np}
\begin{align}\label{Wald field}
A = -\frac{Q}{2M}\eta+\frac{B}{2} (\psi +2 a \eta) \quad ,
\end{align}
where $\eta=\partial_t$ and $\psi =\partial_\phi$ correspond to the Killing vectors associated with the time translation invariance and the axial symmetry of the Kerr geometry respectively. 
The first term is due to the charge of the black hole while the second term is due to an asymptotically uniform magnetic field of magnitude $B$. Notice that we take the charge $Q\ll M $ and magnetic field $B\ll 1/M$ in order to consistently use the Kerr metric, which is a vacuum solution, as is Eq.\ (\ref{Wald field}).

Physical energies should be expressed in terms of the potential difference from infinity, which is equivalent to making a gauge transformation
$A'_\mu=A_\mu- \partial_\mu \alpha$ with
\begin{align}
\alpha =  \left(\frac{Q}{2 M}-a B\right)t.
\end{align}
Explicitly in terms of metric quantities, the resulting vector potential is
\begin{align}
A_t = & -\left (\frac{Q}{2M}-aB  \right )g_{tt}+\frac{ B }{ 2}g_{t\phi} - \left(\frac{Q}{2 M}-a B\right)\nn\\
A_\phi =& -\left (\frac{Q}{2M}-aB\right ) g_{t\phi}+\frac{B }{2}g_{\phi\phi}\nn\\
A_r =&A_\theta =0 \label{Eq:Ag}
\end{align}
where the constant from the gauge choice has been explicitly subtracted in the final term of $A_t$. As $r\to \infty$, the vector potential approaches the asymptotic form
\begin{align}
A_\mu \to \frac{B}{2}(0,0,0, r^2 \sin^2 \theta ),
\end{align}
which is the vector potential for a uniform magnetic field parallel to the $z$-direction with field strength $B$. The magnetic field is aligned (anti-aligned) with the spin of the black hole if $B>0$ ($B<0$). 

Wald observed that it is energetic favorable for the black hole to acquire charge of value $Q_W=2aMB$, which we will call the Wald charge  \cite{Wald:1974np}. A black hole given a reservoir of charged particles will be in the lowest energy state at the Wald charge if spinning in a magnetic field. The black hole is not energetically driven to discharge, contrary to the common assumption. A spinning charged black hole has its own magnetic dipole field and imitates a pulsar as discussed in \cite{Levin:2018mzg}. 

\subsection{Particle dynamics}

We consider a free test particle of mass $\mu>0$ and charge $q$ living in this background, with Lagrangian\footnote{Here we assume the backreaction of the charge particle on the electromagnetic field or the gravitational field is negligible.}
\begin{align}\label{mass Lag}
\mathcal{L}=\frac{\mu}{2}g_{\mu\nu}\dot{x}^\mu\dot{x}^\nu+q A^\mu \dot{x}_\mu
\end{align}
where the dot denotes the derivative with respect to the proper time $\tau$ of the particle. Because of the time translation and axial symmetry of the spacetime, we have two constants of motion
\begin{align}\label{conserved con}
\frac{\partial \mathcal{L}}{\partial \dot{t}}=p_t +q A_t =& - \mu e \nn\\
\frac{\partial \mathcal{L}}{\partial \dot{\phi}} = p_\phi +q A_\phi =& \mu \ell
\end{align}
where $p_\mu = \mu \dot{x}_\mu$ is the kinetic momentum of the particle. The constants $e$ and $\ell$ are, respectively, the conserved energy and the angular momentum in the $z$-direction per unit mass. While it is customary to refer to $\mu e$ and $\mu \ell$ as energy and angular momentum -- and we will continue to do so throughout -- they are not necessarily the energy or angular momentum as measured by any physical observer, as will be relevant for the Penrose process.

Note that with the potential \eqref{Eq:Ag}, a particle at rest at infinity has $e=1$. The only contribution to the energy is its rest mass. Another constant of motion for any timelike geodesic is 
\begin{align}\label{mass}
-\mu^2 = g^{\mu\nu}p_\mu p_\nu,
\end{align}
The full equations of motion that will preserve these 3 constants are summarized by
\begin{align}\label{Eq:EOM}
\left (u\cdot D \right )u =\bar q F\cdot u
\end{align}
with the usual Maxwell tensor $F_{\mu \nu}=\partial_{\mu} A_{\nu }-\partial_{\nu}A_{\mu}$ and charge-to-mass ratio $\bar q=q/\mu$.

As is well known, all particles are dragged around with the spinning spacetime. To find the bounds on the allowed range of a particle's angular velocity, consider a photon emitted at some fixed radial distance $r$ in the $\phi$-direction. At that instant, the angular velocity is
\begin{align}
\Omega_\pm=\bigg(\frac{d\phi}{dt}\bigg)_\pm=-\frac{g_{t\phi}}{g_{\phi\phi}}\pm \sqrt{\bigg( \frac{g_{t\phi}}{g_{\phi\phi}}\bigg)^2-\frac{g_{tt}}{g_{\phi\phi}}}
\end{align}
the $\pm$ correspond to the directions against and along the rotation of the black hole respectively. For any massive particle, its angular velocity is then bounded by
\begin{align}
\Omega_- \leq \Omega \equiv \frac{d\phi}{dt} \leq \Omega_+.
\end{align}
Using the metric relationships, we have
\begin{align}\label{null ang}
\Omega_\pm=-\frac{g_{t\phi}}{g_{\phi\phi}}\pm \sqrt{\frac{\Delta \sin^2\theta}{g_{\phi\phi}^2}}
\end{align}
which makes clear that on the outer horizon $r_+$, the square root in \eqref{null ang} becomes zero and we have
\begin{align}\label{min ang}
\Omega_H = \left.\frac{d\phi}{dt} \right |_{r=r_+}
=-\left.\frac{g_{t\phi}}{g_{\phi\phi}}\right |_{r=r_+}
=\frac{a}{r_+^2+a^2}.
\end{align}
This is the minimum angular velocity of a particle at the horizon due to the frame dragging effect. 

An interesting special observer to consult is the ZAMO (zero angular momentum observer), whose worldline has $u_Z^\mu$ and $\psi \cdot u_Z=\ell=0$. 
Solving for the angular velocity gives
\begin{align}
 \Omega_Z(r)= \frac{d\phi}{dt}=-\frac{g_{t\phi}}{g_{\phi\phi}}>0,\quad \text{for} \quad r>r_+ \quad .
\end{align}
The 4-velocity of the ZAMO is then
\begin{align}
u_Z = u_Z^t (\eta +\Omega_Z \psi) \ \ ,
\end{align}
with $u_Z^t>0 $ as set by $u\cdot u=-1$.

 As $r$ increases, $\Omega_Z$ is monotonically decreasing. In particular this means
\begin{align}
\Omega_H\geq \Omega_Z ,\quad r\geq r_+
\end{align}
and the equality holds on the horizon.

The mechanical Penrose process cleverly exploits the relativity of space and time to augment the energy of outgoing particles. Consider a particle 1 that decays into particle 2 and particle 3.
Simply put, conservation of 4-momentum enforces
\begin{equation}
    \mu_1 e_1=\mu_2 e_2+\mu_3 e_3 \quad .
\end{equation}
Particle 3 can emerge with more energy than its parent if $e_2<0$. While a negative kinetic energy is impossible, $e_2<0$ is not necessarily the energy as measured by any observer, despite its name. The observed energy of any particle is relative and quantified by its momentum through some observer's time. Since all observers are free to consider themselves to be at rest in their own frames, their time direction is equivalent to their 4-velocity $u$. The requirement that the kinetic energy of a particle with momentum $p$ be positive as measured by our local observer is $-p\cdot u\ge 0$. As long as this condition is respected for all viable, local observers, no laws have been broken.
Within the ergosphere, there is no observer who can naively interpret $\mu e$ as kinetic energy. Indeed, the $t$-component of momentum is interpreted as a spatial momentum and $e$'s positivity is no longer enforced. 
Consequently, an outgoing particle can have $\mu_3 e_3$ greater than $\mu_1 e_1$ of the original parent if the other daughter has a negative $\mu_2 e_2$, as we explicitly demonstrate in Sec. \ref{Sec:decay}. 

The lesson for now is that we are in search of  negative $e$ values for one daughter in order for the other to extract energy. In the electromagnetic Penrose process, the negative $e$ regions are no longer strictly set by the ergosphere. We set out to find the negative energy states and the extended Penrose regions in Sec. \ref{Sec:EnCon}.

\subsection{ Electromagnetic couplings}
\label{Sec:EMCouplings}

Before we identify negative energy states, we
streamline notation with the introduction of the following dimensionless parameters
\begin{align}\label{para}
\chi_Q = \qs \ , \quad\quad
\chi_B =\bar{q} \frac{BM}{2}
\end{align}
to reexpress the vector potential as 
\begin{align}
\bar q A_t = & -\chi_Q \left (g_{tt}+1\right )+\frac{\chi_B}{M} g_{t\phi}\nn\\
\bar q A_\phi =& -\chi_Q g_{t\phi}+\frac{\chi_B}{M} g_{\phi\phi}\quad ,\label{Eq:Achi}
\end{align}
which reveals that $\chi_Q$ and $\chi_B$ capture the electromagnetic couplings between the black hole, the charged particle, and the background magnetic field.

There are 4 possible sign combinations of $\chi_B$ and $\chi_Q$, which depend on (i) whether the magnetic field is aligned or antialigned with the black hole spin, (ii) the charge of the particle, and (iii) whether the black hole charge $Q$ has exceeded the Wald value $Q_W=2aMB$. When the magnetic field is aligned with the black hole spin, i.e. $B>0$, we have $Q_W>0$ and 
\begin{center}
\begin{tabular}{ |c|c|c| } 
 \hline
  & $\chi_B>0$ & $\chi_B<0$ \\ 
 \hline
 $\chi_Q>0$ & $\bar q>0$ and $Q>Q_W>0$ & $\bar q<0$ and $Q<Q_W$ \\ 
 \hline
 $\chi_Q<0$ & $\bar q>0$ and $Q<Q_W$ & $\bar q<0$ and $Q>Q_W>0$ \\ 
 \hline
\end{tabular}
\end{center}
while when $B<0$ we have $Q_W<0$ and 
\begin{center}
\begin{tabular}{ |c|c|c| } 
 \hline
  & $\chi_B>0$ & $\chi_B<0$ \\ 
 \hline
 $\chi_Q>0$ & $\bar q<0$ and $Q<Q_W<0$ & $\bar q>0$ and $Q>Q_W$ \\ 
 \hline
 $\chi_Q<0$ & $\bar q<0$ and $Q>Q_W$ & $\bar q>0$ and $Q<Q_W<0$  \\ 
 \hline
\end{tabular}
\end{center}

For subatomic particles, the charge-to-mass ratios are typically very large, which makes $\chi_B$ typically large in realistic situations:
\begin{align}
\chi_B \approx 2\times 10^{15}\left( \frac{\bar{q}}{\bar{q}_p}\right) \left(\frac{M}{10M_\odot}\right) \left(\frac{B}{10^{12}G}\right).
\end{align}
Here $\bar{q}_p \approx 10^8 \text{C} \cdot \text{kg}^{-1}$ is  the charge-to-mass ratio for a proton. In a binary system where the uncharged black hole with $M=10 M_\odot$ is aligned with a uniform magnetic field $B\approx 10^{12} \text{ to }10^{15} G$ created by a neutron star \cite{NeturonStarMag},  a proton has $\chi_B \approx 10^{15}\text{ to }10^{18}$ while an electron has $\chi_B \approx -10^{18}\text{ to }-10^{21}$. Another example is an uncharged supermassive black hole like M87 with $M\approx 6.5 \times 10^6 M_{\odot}$ immersed in a magnetic field $B \approx 30 G$ \cite{MagFieldEHT}, in which case a proton has $\chi_B \approx 10^{10}$ and an electron has $\chi_B \approx -10^{13} $. Similarly, for an uncharged black hole, we have
\begin{align}
\label{Eq:chiQbig}
\chi_Q \approx -4\times10^{15}\left( \frac{\bar{q}}{\bar{q}_p}\right)\left(\frac{a}{M}\right) \left(\frac{M}{10M_\odot}\right)\left(\frac{B}{10^{12}G}\right) .
\end{align}
The Wald mechanism \cite{Wald:1974np} energetically favors a spinning black hole acquire charge in a magnetic field, so that the magnitude of $\chi_Q$ decreases. Energetically, the black hole is disinclined to charge beyond the Wald value $Q_W=2aBM$ ($\chi_Q=0$). Assuming there are no other competing mechanisms that charge or discharge the black hole, we argue that the realistic range of $\chi_Q$ is
\begin{equation}
    -2 \frac{a}{M} \chi_B < \chi_Q <0
    \label{chiQneg}
\end{equation}  
if $\chi_B >0$ 
and
\begin{equation}
    -2 \frac{a}{M} \chi_B > \chi_Q >0
\end{equation}
if $\chi_B<0$. 

As we show in the next section, the electromagnetic energy boost is more powerful for $\chi_Q<0$, so the optimum range is \eqref{chiQneg}, for which
$\chi_B>0$ and $\chi_Q<0$. This is equivalent to a range extending from uncharged black holes, for which the superradiance will be largest, to black holes that reach the Wald charge, for which the superradiance will be smallest. We will continue to consider general ranges as indicated.

Hereafter we work in natural units with $M=1$.

\section{ENERGY CONDITIONS}
\label{Sec:EnCon}

\subsection{Minimum energy}

We can require of all particles that a ZAMO at a location $r>r_+$ sees a particle to have positive kinetic energy. In other words, 
\begin{align}
p\cdot u_Z<0 \ \ .
\end{align}
This gives a condition on the energy of the particle, $e \ge e_{min}$, with 
\begin{align}\label{Z bound}
e_{min} = &  \Omega_Z  (\ell-\bar{q} A_\phi) - \bar{q}A_t \nn \\
= &  \Omega_Z  \ell +\left (1-\frac {\Delta \sin^2\theta}{g_{\phi\phi} }\right )\chi_Q
\quad . \end{align}
Notice this is true for all $(r,\theta)$. 
The quantity $e_{min}$ can be negative, even when $\ell\ge 0$, and still have a physically meaningful positive kinetic energy. While the ZAMO requires $e>e_{min}$,
we can do better and find exactly how much bigger $e$ is than $e_{min}$ below in Eq.\ \eqref{Eq:elong}.

We could require that the observed particle with $e<0$ crosses the event horizon, which for a stationary spacetime is a Killing horizon. For the Kerr case, one can show that the following linear combination of time-translation and rotational Killing vectors
\begin{align}\label{hor gen}
\xi = \eta + \Omega_H \psi
\end{align}
generates the horizon. On $r=r_+$, $\xi$ becomes null. The condition that a particle crosses the event horizon moving forward in time is
\begin{align}
p^\mu \xi_\mu <0.
\end{align}
We then have the condition $e>e_{min}(r_+)$
\begin{align}\label{gen bound}
e_{min}(r_+)= & \Omega_H  (\ell-\bar{q} A_\phi) - \bar{q}A_t \nn \\
 =  & { \Omega_H  \ell  +\chi_Q }
\end{align}
where the last line is obtained using Eq.\ (\ref{Eq:Ag}) and the metric relations. Eq. (\ref{gen bound}) matches Eq. (\ref{Z bound}) at the horizon, as it must. 

Unlike the Kerr case, when $e$ becomes negative, $\ell$ does not have to be negative. The permitted parameter ranges for $e$ and $\ell$ depend on the vector potential $A_\mu$ and the location of the particle in a detailed way. Physically, this is sensible. One can imagine extracting energy from the black hole electromagnetically instead of mechanically extracting rotational energy.

\subsection{Negative energy states}

We are now at the crux, which is to remap the ergosphere to a new negative energy region. Above we found the minimum values for $e$ that lead to a positive energy as measured by a ZAMO in Eq.\ (\ref{Z bound}), with the special condition that the observed particle crosses the horizon in Eq.\ (\ref{gen bound}). These are rock bottom values of $e$ for physical plausibility. But we actually know $e$ in terms of other variables and can explore if $e$ ever probes the range $e
_{min}<e<0$.

Using the constants of the motion from the previous section, we solve for $e$. From the timelike constraint (\ref{mass}), we can express $e$ in terms of other quantities. Eliminating $p_t$ and $p_\phi$ gives
\begin{align}
 e^2 -2 \beta e +\gamma=0 
\end{align}
where
\begin{align}\label{E eq para}
\beta= & - \bar{q}A_t +\Omega_Z(\ell-\bar{q} A_\phi)\nn\\
\gamma=&\bar{q}A_t \Big(\bar{q}A_t -2\Omega_Z(\ell-\bar{q} A_\phi)\Big)+  \frac{g_{tt}}{g_{\phi \phi}}  (\ell-\bar{q} A_\phi)^2 -\frac{ \Delta \sin^2\theta}{g_{\phi\phi}} \bigg(\frac{g^{rr}p_r^2+g^{\theta\theta}p_\theta^2}{\mu^2}+1 \bigg).
\end{align}
giving
\begin{align}
\label{Eq:elong}
e=&\beta + \sqrt{\beta^2-\gamma}\nn\\
=& e_{min} +\bigg[\frac{ \Delta \sin^2\theta}{g_{\phi\phi}}\bigg(\frac{(\ell-\bar{q} \frac{B }{2}g_{\phi\phi})^2}{g_{\phi\phi}}+\frac{g^{rr}p_r^2+g^{\theta\theta}p_\theta^2}{\mu^2}+1 \bigg)\bigg]^{1/2} \quad
\end{align}
where $e_{min}$ is defined in Eq.\ \eqref{Z bound}.
We can express the energy $e$ per unit mass when $p_r=p_\theta=0$ as 
\begin{align}\label{energy}
e_*
=& e_{min}+\bigg[\frac{ \Delta \sin^2\theta}{g_{\phi\phi}}\bigg(\frac{(\ell-\bar{q} A_\phi)^2}{g_{\phi\phi}}+1 \bigg)\bigg]^{1/2}.
\end{align}
As $\ell\rightarrow -\infty$, $e_*$ approaches its value for the purely mechanical Penrose process. The sign of the radical is chosen so that it corresponds to a particle moving forward in time with respect to a ZAMO. The fact that $e$ can be negative is what allows extraction of energy. The mechanical Penrose process corresponds to $A_t =A_\phi =0$, for which $e<0$ is possible only within the ergosphere. The presence of a nonzero vector potential enables a magnetic Penrose process that allows for electromagnetic energy extraction. Note that the $-\bar{q} A_t $ term in \eqref{energy} potentially extends the region of negative energy orbits all the way to infinity \cite{PhysRevD.29.2712,PhysRevD.30.1625}. In other words, if we do not restrict the charge of the black hole, we can always mine electromagnetic energy from anywhere. However, if we restrict the charge to the energetically favored range, $0\le |Q|\le |Q_W|$, then the regions from which electromagnetic energy can be mined are restricted. 

For particles that cross the event horizon, 
\begin{align}
\label{Eq:nofall}
e=e_{min}(r_+)+\frac{r^2_++a^2\cos^2\theta}{2Mr_+}|\dot r|_{r=r_+}
=\Omega_H\ell+\chi_Q+\frac{r^2_++a^2\cos^2\theta}{2Mr_+}|\dot r|_{r=r_+}
\end{align} 
where the final term is always greater than or equal to zero. If $e<e_{min}(r_+)$, then that particle cannot fall in to the black hole. It will either orbit or escape.

From \eqref{Eq:nofall}, we can see that
the scale of the negative electromagnetic energies that can be attained can be estimated by $e_{min}(r_+,\ell=0,\dot r=0)=\chi_Q$. Therefore, $\chi_Q>0$ suppresses energy extraction, and the energy states with largest negative values correspond to $\chi_Q\ll 0$. In a decay process, for instance, the positive-energy particle will then get a boost of energy above the parent on order $-\chi_Q > 0$. The natural magnitude of $\chi_Q$ therefore sets the magnitude of the energy output expected. As shown in Eq.\ \eqref{Eq:chiQbig}, $\left |\chi_Q\right |\gg 1$ and therefore the energy from the electromagnetic Penrose process can be very large, much larger than the $\sim 20\%$ boost of the mechanical process.

Notice that if $a$ and $B$ are aligned, $\chi_Q$ is only negative below the Wald charge for positive charges and is only negative above the Wald charge for negative charges. In other words, the black hole tends to charge up in the Penrose process until it reaches the Wald charge and tends to discharge above the Wald charge. 

Hereafter, we work with moderate values of $\chi_B$ and $\chi_Q$ for ease of computation and discuss the qualitative effects of increasing their magnitudes where appropriate.

\subsection{The zero-energy surfaces}
\label{Sec:ZES}

The ergosphere is defined as the region bounded by the surface with $g_{tt}=0$ given by $r_e(\theta)=1 + \sqrt{1 -a^2\cos^2\theta}$. In the mechanical Penrose process, all negative energy states occur within the ergosphere. The stationary surface bounding the ergosphere coincides with the largest zero-energy surface defined by $e_*(r)=0$, in the absence of electromagnetic fields.

For the electromagnetic Penrose process, 
the largest zero-energy surfaces are given by the $r(\theta)$ for which $e_*=0$ in Eq.\ (\ref{energy}). The boundary of the ergosphere and the zero-energy surface will always coincide when $\ell\rightarrow -\infty$. In the mechanical Penrose process with less negative $\ell$, the zero-energy surfaces are smaller than the ergosphere as in Fig.\ \ref{ZEmech}. In the following figures, we project in spatial coordinates, which correspond to the oblate spheroidal coordinates when we take the $M\to 0$ limit of the Kerr background, using the transformation
\begin{align}
    x = \sqrt{r^2+a^2}\sin\theta\cos\phi ,\qquad y = \sqrt{r^2+a^2}\sin\theta\sin\phi ,\qquad z = r\cos\theta.
\end{align}
\begin{figure}[H]
\centering
  \includegraphics[scale=0.6]{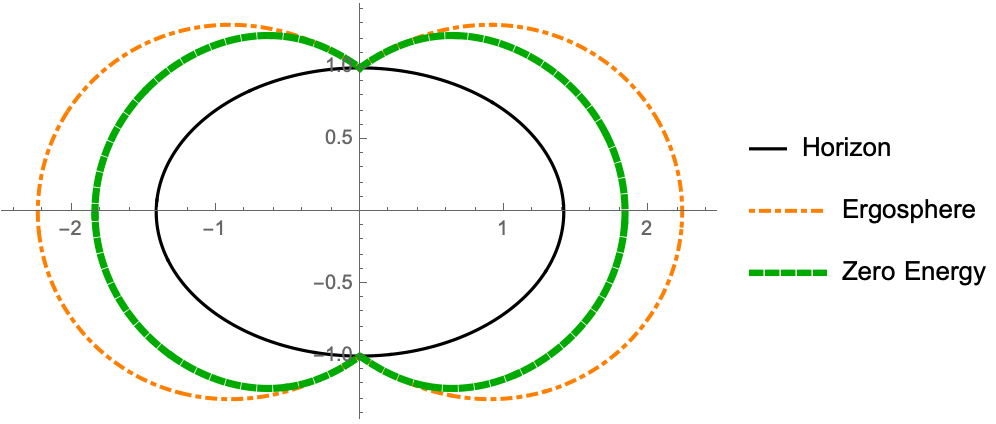}
 \caption{Mechanical zero-energy surface for $\ell=-1$ ($\chi_Q=\chi_B=0$) for $y=0$. As $\ell$ gets more negative, the zero-energy surface approaches the boundary of the ergosphere.}
 \label{ZEmech}
\end{figure}
To proceed, we separate two cases according to the sign of the product $\chi_B\chi_Q$.

\subsubsection{$\chi_B \chi_Q<0$}

\begin{figure}[H]
  \minipage{0.40\textwidth}
  \qquad
 \includegraphics[width=\linewidth]{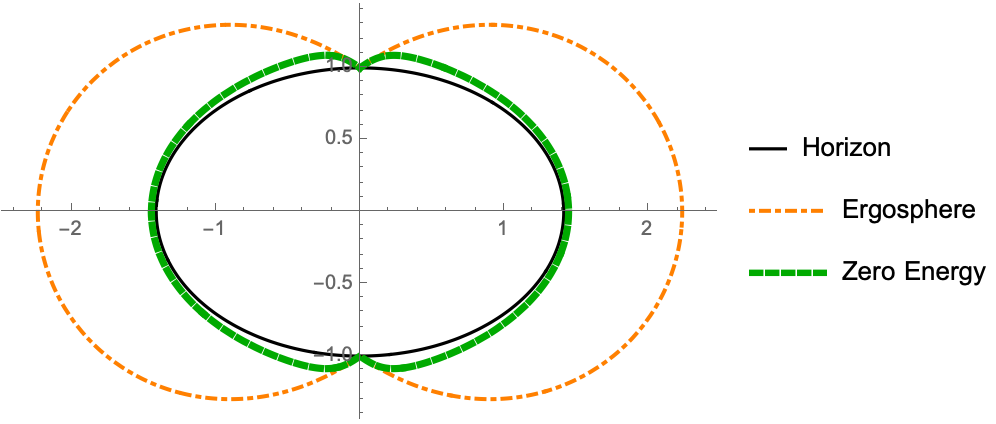}
\endminipage 
\minipage{0.20\textwidth}
\hspace{0.20\textwidth}
\endminipage
  \minipage{0.30\textwidth}
   \includegraphics[width=\linewidth, height = 2\linewidth]{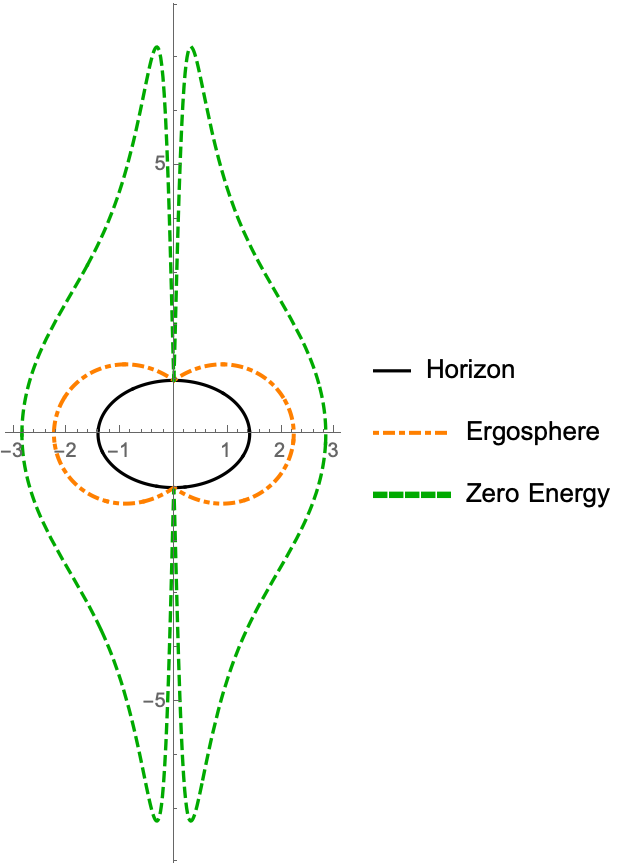}
  \endminipage\hfill
   \minipage{0.40\textwidth}
   \qquad
   \includegraphics[width=\linewidth]{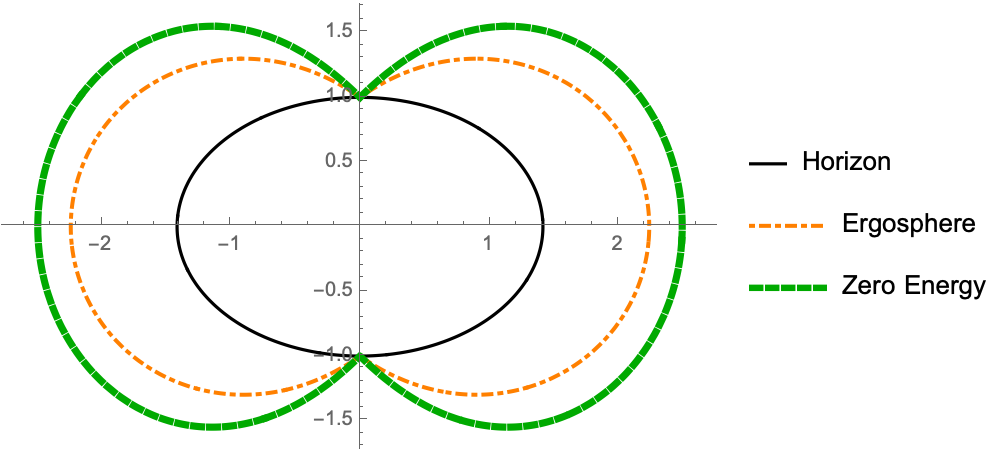}
  \endminipage \qquad\qquad\qquad
  \minipage{0.40\textwidth}
  \includegraphics[width=\linewidth]{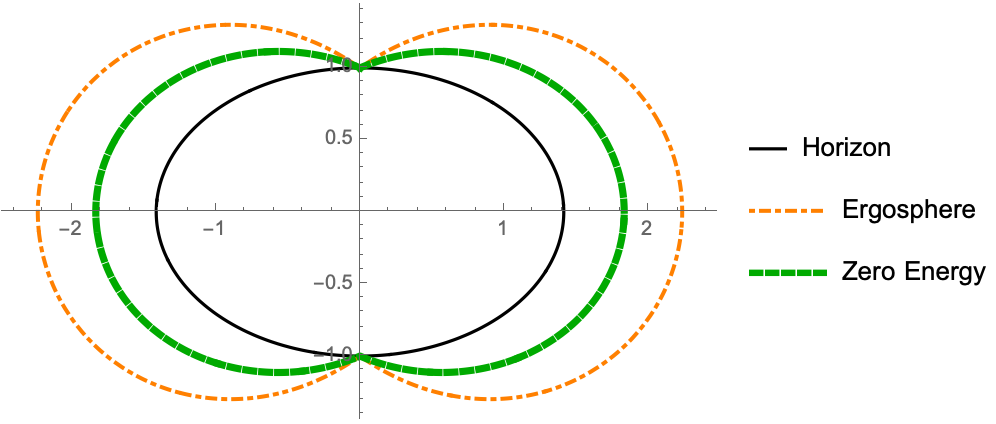}
  \endminipage 
  \caption{In all images, the solid line is the event horizon, the dot-dashed line is the boundary of the ergosphere, and the thick dashed line is the zero-energy surface. In all cases, as $\ell$ gets more negative, the zero-energy surface approaches the ergosphere. Upper left : a black hole at the Wald charge ($\chi_Q=0,\chi_B=10,\ell=-1$). The larger $B$, the smaller the zero-energy surface. Upper right: an uncharged black hole ($\chi_Q=-2a \chi_B,\chi_B=10,\ell=-1$). Notice that as compared with the other figures at the same $\ell$, the zero-energy surface exceeds the ergosphere. Lower left: same as upper right but with $\ell=-100$. Lower right: same as lower left, but with $\chi_B=-10$.}
  \label{ZEScompare}
\end{figure}

As discussed in at the end of the last section, this sign combination is of realistic interest. For the enhanced case with $\chi_B> 0$ and $\chi_Q<0$, the zero-energy surface can be larger than the ergosphere when the black hole is uncharged as in the upper right image in Fig.\ \ref{ZEScompare}, as opposed to the suppressed case . The larger $\chi_B$, the zero-energy surface elongates near the poles, scaling roughly as $z\propto \chi_B$. By contrast, if we restrict to $-2 \chi_B\leq \chi_Q<0$, the zero energy surface cannot extend too far from the horizon in the equatorial plane. In fact, it is not difficult to show that for $\chi_Q =-2 \chi_B$ (chargeless black hole), one can estimate the location on the equatorial plane where $e_*(r)=0$ to be (for large $\chi_B$)
\begin{align}
r\approx \frac{1}{3} \left(2+\sqrt[3]{44-3 \sqrt{177}}+\sqrt[3]{44+3 \sqrt{177}}\right) \approx 2.65897,
\end{align}
compared to the location of the stationary surface at $r=2$. As $\ell$ gets very negative for fixed $\chi_Q,\chi_B$, the zero-energy surfaces approach the ergosphere, as in the lower left image. At the same $\ell=-100$, we show the suppressed case in the lower right image.

When the black hole has attained the Wald charge ($\chi_Q=0$), the zero-energy surface is smaller than the ergosphere as in the upper left image in Fig. \ref{ZEScompare}. The larger $\chi_B$, the smaller the zero-energy surface.

\subsubsection{$\chi_B \chi_Q>0$ }

As discussed in at the end of the last section, this case is unlikely occurring in nature, but we will give some brief comments for theoretical interest.

\begin{figure}[H]
\minipage{0.07\textwidth}
  \hspace{0.07\textwidth}
  \endminipage
\minipage{0.3\textwidth}
  \includegraphics[width=1.3\linewidth]{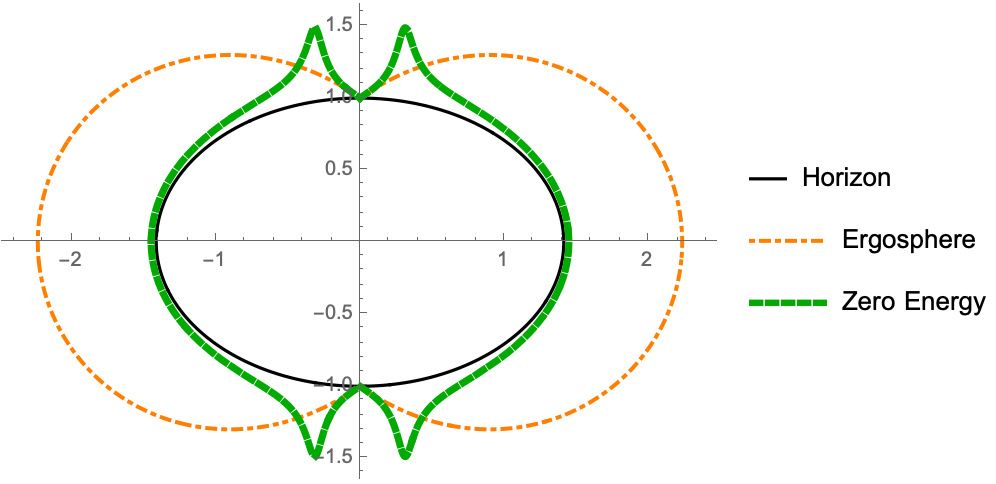}
  \endminipage\qquad \qquad \qquad \qquad
  \minipage{0.1\textwidth}
  \hspace{0.1\textwidth}
  \endminipage
  \minipage{0.2\textwidth}
 \includegraphics[width=\linewidth]{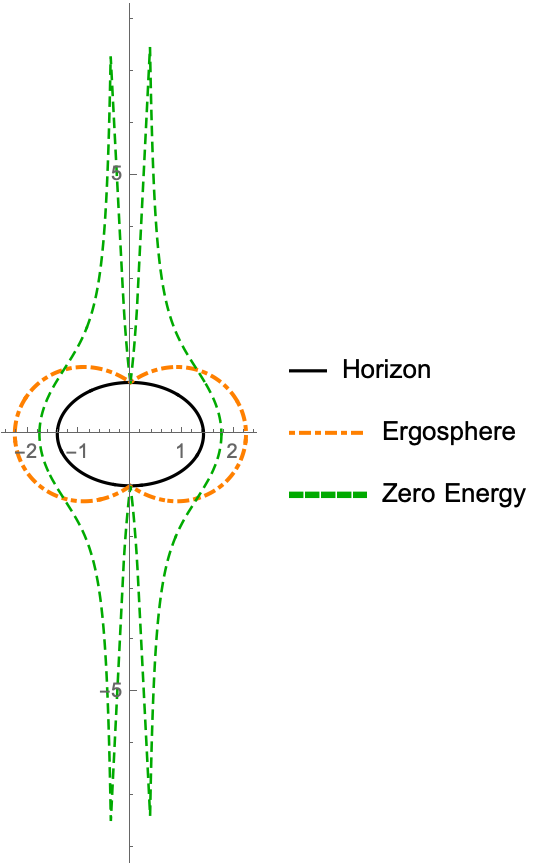}
\endminipage\hfill
\minipage{0.13\textwidth}
  \hspace{0.13\textwidth}
  \endminipage
  \minipage{0.25\textwidth}
   \includegraphics[width=\linewidth]{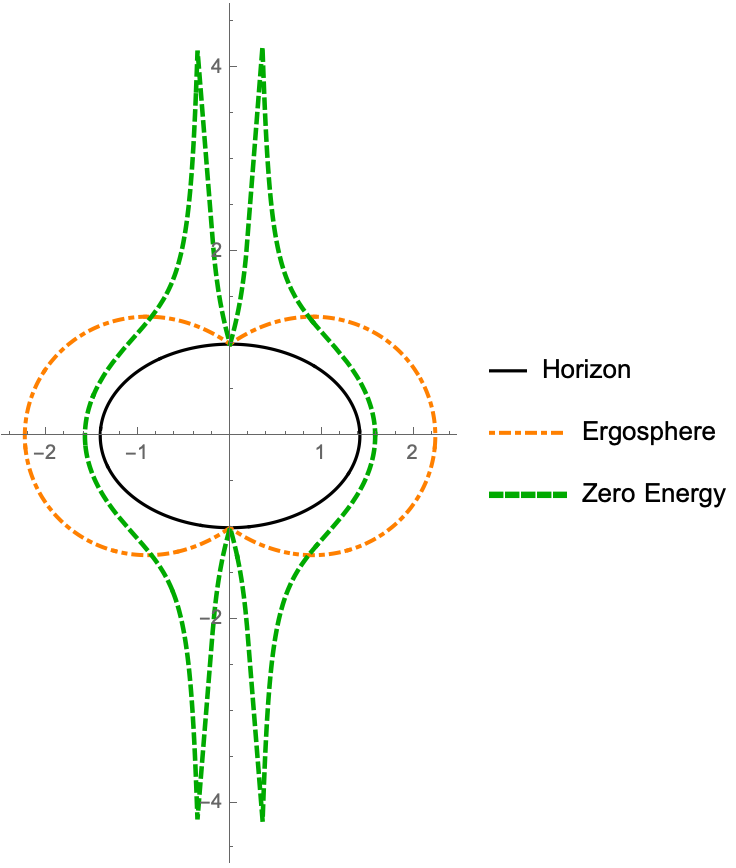}
  \endminipage\qquad \qquad \qquad
   \minipage{0.3\textwidth}
   \includegraphics[width=1.3\linewidth]{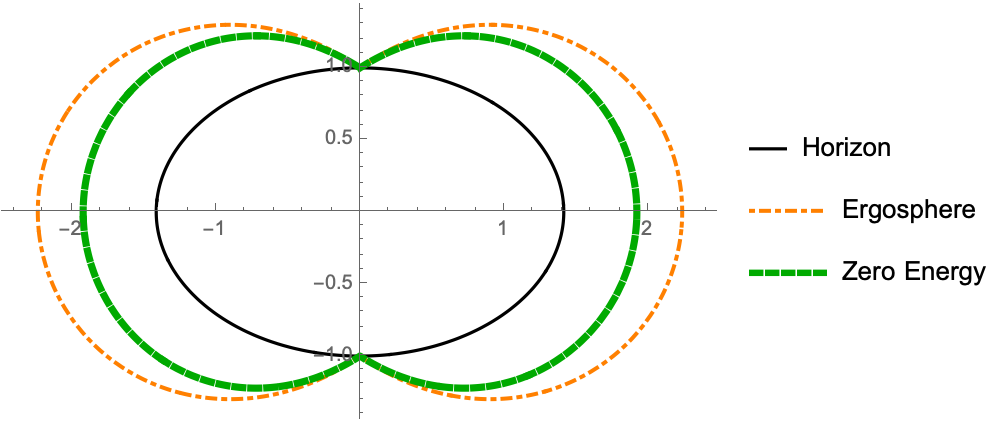}
  \endminipage\hfill
  \caption{Upper Left: 
  ($\chi_Q=0,\chi_B=-10,\ell=-1$). 
  Upper Right:  ($\chi_Q=-5,\chi_B=-10,\ell=-1$). Lower Left: ($\chi_Q=-2,\chi_B=-10,\ell=-1$). Lower Right:  ($\chi_Q=5,\chi_B=10,\ell=-100$).}
  \label{antiZEScompare}
\end{figure}

We first comment on the case with $\chi_B<0$ and $ \chi_Q<0$, where we have a suppression of the Penrose process. The upper left image of Fig. \ref{antiZEScompare}, shows the zero-energy surface at the Wald charge. When the black hole charges up beyond the Wald charge, the zero-energy surface will expand and elongate near the poles, as shown in the upper right and lower left images. Finally, an example for suppression with $\chi_B>0$ and $ \chi_Q>0$ is shown in the lower right image, which should be compared with the lower left image in Fig. \ref{ZEScompare}.


\subsection{Toroidal zero-energy surfaces}
\label{Sec:multiple}

An intriguing feature of the black hole immersed in a magnetic field is that there can be toroidal zero-energy surfaces that are not coincident with the ergosphere. To our knowledge, this is the first demonstration of toroidal negative-energy regions, which furthermore do not contain the event horizon. 

Notice that if we restrict to initial values with $p_r=p_\theta=0$ so that $e=e_*$, the zero-energy surfaces tell us about the stability of the orbits.
In all cases, negative energy particles with $e=e_*<0$ are trapped in the zero-energy surfaces and therefore never escape to infinity. If they live within a surface that includes the event horizon, they will presumably fall into the black hole. If they are within a toroidal region, they can only be on stable orbits that never reach the event horizon.

To search for multiple negative-energy regions, we search for multiple zeroes of $e_*(r)$.
Keeping $a=1$ for simplicity, we recall the definition \eqref{energy} in terms of \eqref{E eq para}. As in the analysis in \cite{1985JApA....6...85B}, finding the zeros of $e_*(r)$ is equivalent to finding the locations where 
\begin{align}
\gamma =0 \qquad \text{and} \qquad \beta<0.
\end{align}
The $\beta<0$ constraint reads simply
\begin{align}\label{b ineq}
 \ell + \chi_Q (1+(1+R)^2)<0.
\end{align}
To proceed, we focus on the $\gamma =0$ condition. Restricting ourselves on the equatorial plane $\theta=\frac{\pi}{2}$ and plugging in the explicit expressions for the vector potentials, this is equivalent to
\begin{align}
\Big[ \chi_Q (\Delta-g_{\phi\phi})+g_{t\phi}\ell\Big]^2-\Delta\Big[ \left( \chi_Q g_{t\phi} -\chi_Bg_{\phi\phi}+\ell \right)^2-g_{\phi\phi}\Big]=0.
\end{align}
We want to examine the solution to this equation outside the horizon, i.e. $r>r_+=1$. To that end, we change the variable $R=r-1>0$ and expand the left-hand side. This is then equivalent to the 5-order polynomial
\begin{align}\label{poly}
0 = R^5+3 R^4 + c_3 R^3+c_2 R^2+c_1 R+c_0
\end{align}
where
\begin{align}
c_3=-\frac{2 \ell}{\chi_B}+\frac{1}{\chi_B^2}+4,\quad c_2=&  \frac{4 \chi_Q-2 \ell}{\chi_B}+\frac{1}{\chi_B^2}+4, \quad c_1=&\frac{\ell^2-4 \chi_Q^2}{\chi_B^2}, \quad c_0 = -\frac{(\ell+2 \chi_Q)^2}{\chi_B^2}.
\end{align}
The positive zeros to the Eq. \eqref{poly} obeying $\beta<0$ correspond to zero-energy surfaces. If there is only one such root, there is only one such surface which together with the outer horizon bounds a region where $e_*(r)$ is negative. If there are multiple such solutions to \eqref{poly}, then there can be multiple surfaces with zero energy. To examine this problem, we employ Descartes' rule of signs\footnote{Descartes' rule of signs says that the number of sign changes in the sequence of a polynomial's coefficients (omitting the zero coefficients) is greater than or equal to the number of positive roots. The difference between the number of sign changes and the number of positive roots is always even.}. What we are aiming for is as many sign changes in the sequence $\{c_i\}$ as possible. Since $c_0<0$, there are only 4 possibilities with more than 1 sign change:
\begin{align}\label{possibilities}
\text{(i) } &\quad c_3>0,\quad c_2<0, \quad c_1>0 \nn\\
\text{(ii) } &\quad c_3<0, \quad c_2>0, \quad c_1>0\nn\\
\text{(iii) } &\quad c_3<0,\quad c_2>0, \quad c_1<0\nn\\
\text{(iv) } &\quad c_3<0, \quad c_2<0, \quad c_1>0
\end{align}
To continue the analysis we separate two cases according to the sign of the product $\chi_B\chi_Q$.

\subsubsection{$\chi_B \chi_Q<0$}

In this case, we must have $c_2<c_3$ and thus the possibilities (ii) and (iii) are immediately ruled out. Therefore we must have $c_1>0$ and $c_2<0$. The former implies $\chi_B\ell>-2\chi_B\chi_Q>0$ or $\chi_B\ell<2\chi_B\chi_Q <0$. However, if $\chi_B\ell<2\chi_B\chi_Q <0$, we will have $c_3>c_2>0$, which is not one of the 4 possibilities \eqref{possibilities}. We are then left with 
\begin{align}\label{c1 ineq}
\chi_B\ell>-2\chi_B\chi_Q>0.
\end{align}
The $c_2<0$ condition is equivalent to
\begin{align}\label{c2 ineq}
2\chi_B^2 +\frac{1}{2}<\chi_B(\ell-2\chi_Q).
\end{align}
According to Descartes' rule of signs, if inequalities \eqref{c1 ineq} and \eqref{c2 ineq} are satisfied, the equation \eqref{poly} can have 1 or 3 positive roots. However, even if \eqref{poly} has 3 positive roots, it does not mean that there are 3 surfaces of zero energy, because we still have the condition \eqref{b ineq}. Only the satisfaction of all  \eqref{c1 ineq}, \eqref{c2 ineq} and \eqref{b ineq} can possibly lead to multiple zero-energy surfaces.

\begin{figure}[H]
\centering
    \begin{subfigure}[b]{0.5\textwidth}
        \raisebox{0.4\height}{\includegraphics[width=\textwidth]{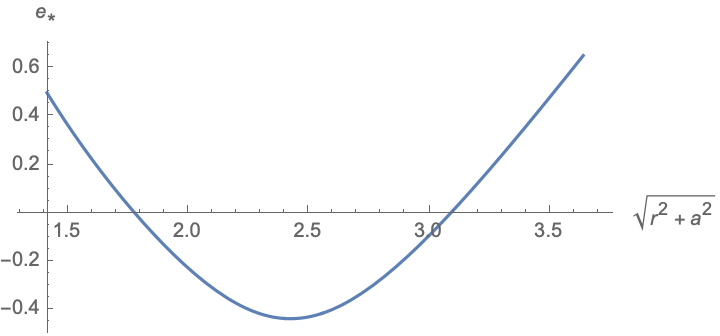}}
    \end{subfigure}
    \qquad 
    \begin{subfigure}[b]{0.4\textwidth}
        \includegraphics[width=\textwidth]{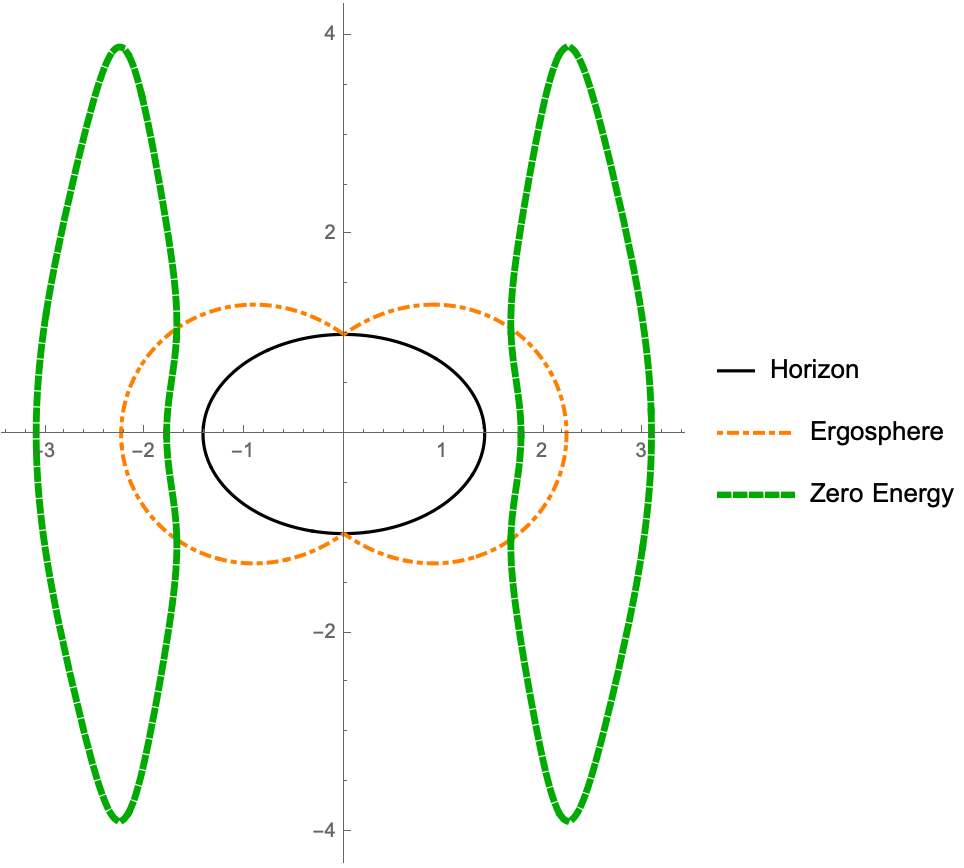}
    \end{subfigure}   
  \caption{The case with $a=1,\chi_B = 1,\chi_Q =-2 ,\ell=5$. In this case $e_*(r)$ has 2 zeros outside the horizon as shown in the left figure for $\theta=\pi/2$. On the right, a toroidal zero-energy surface is shown for an uncharged black hole. }
  \label{2zeros}
\end{figure}

\begin{figure}[H]
\centering
    \begin{subfigure}[b]{0.5\textwidth}
        \raisebox{0.3\height}{\includegraphics[width=\textwidth]{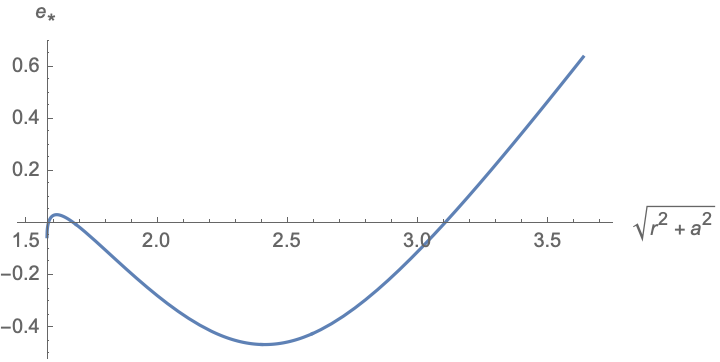}}
    \end{subfigure}
    \qquad 
    \begin{subfigure}[b]{0.4\textwidth}
        \includegraphics[width=\textwidth]{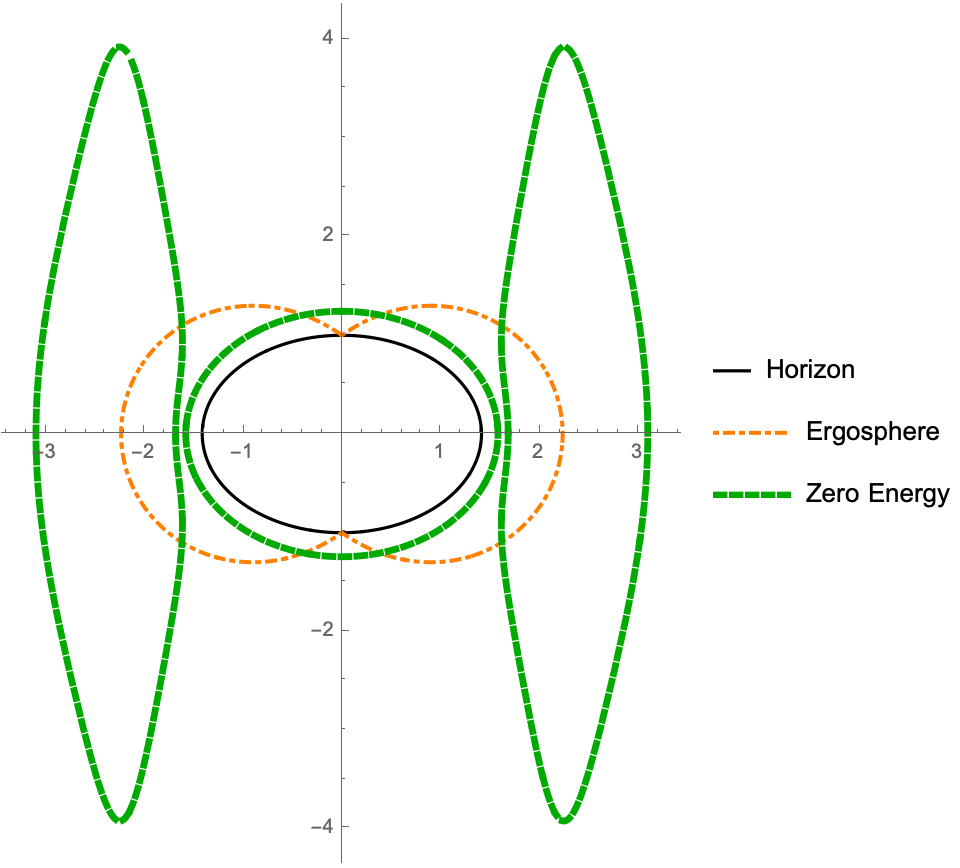}
    \end{subfigure}   
  \caption{The case with $a=0.97,\chi_B = 1,\chi_Q =-2 ,\ell=5$. In this case $e_*(r)$ has 3 zeros outside the horizon as shown in the left figure for $\theta=\pi/2$. On the right, various surfaces are shown. Note that the surface that is closest to the horizon almost coincides with the horizon.}
  \label{3zeros}
\end{figure}

\begin{figure}[H]
\centering
    \begin{subfigure}[b]{0.4\textwidth}
        \includegraphics[width=\textwidth]{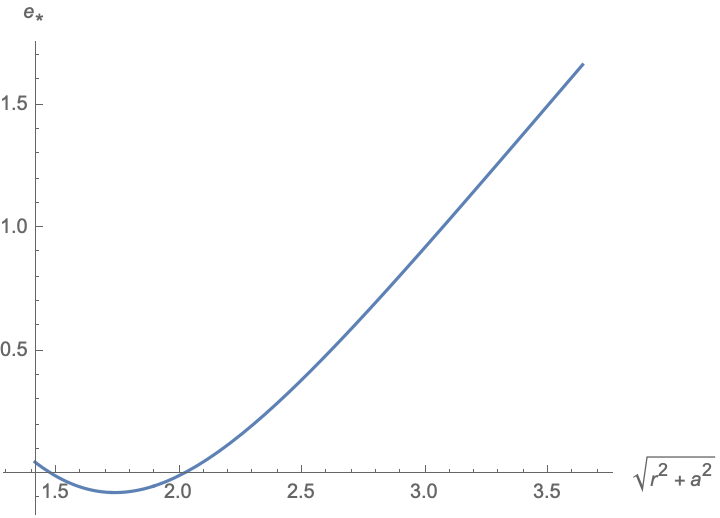}
    \end{subfigure}
    \quad 
    \begin{subfigure}[b]{0.45\textwidth}
        \includegraphics[width=\textwidth]{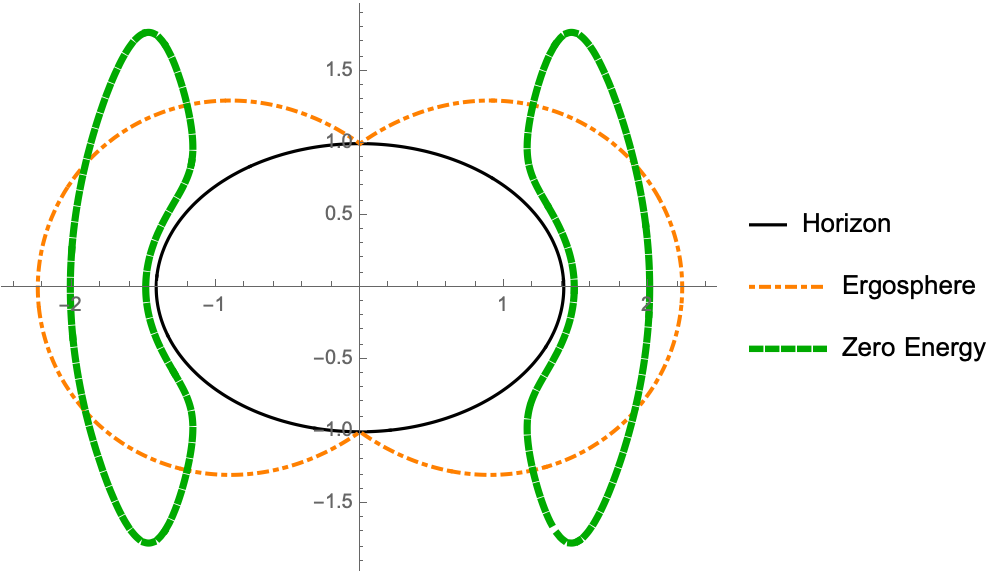}
    \end{subfigure}   
  \caption{The case with $a=1,\chi_B = 1,\chi_Q =-1 ,\ell=2.1$. In this case $e_*(r)$ has 2 zeros outside the horizon as shown in the left figure for $\theta=\pi/2$. On the right, a toroidal zero-energy surface moves toward the horizon as the black hole charges up.}
  \label{2zeros charged}
\end{figure}

\begin{figure}[H]
\centering
    \begin{subfigure}[b]{0.3\textwidth}
        \includegraphics[width=1.1\textwidth]{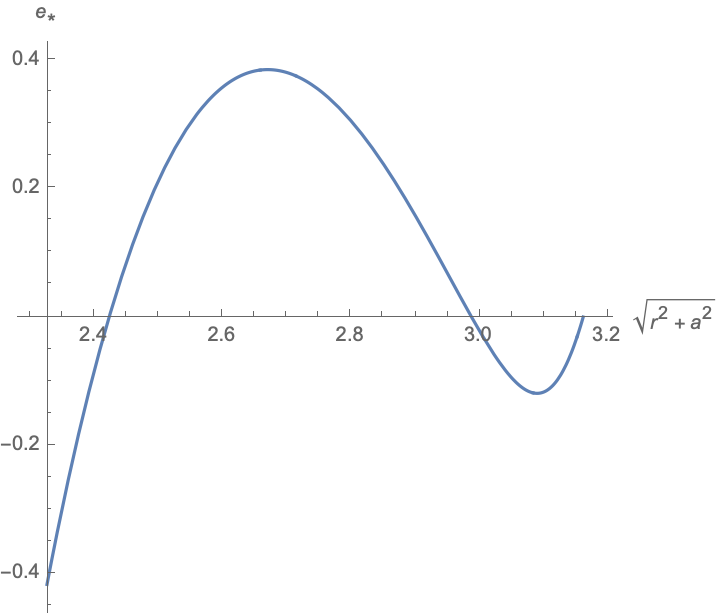}
    \end{subfigure}
    \quad \qquad 
    \begin{subfigure}[b]{0.5\textwidth}
        \raisebox{0.2\height}{\includegraphics[width = 1.2\textwidth]{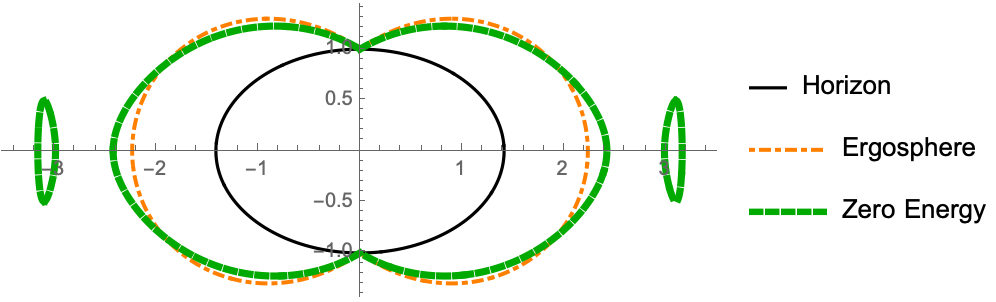}}
    \end{subfigure} 
    \qquad \quad
  \caption{The case with $a=1,\chi_B = -5,\chi_Q =4 ,\ell=-50$. In this case $e_*(r)$ has 3 zeros outside the horizon as shown in the left figure for $\theta=\pi/2$. On the right, zero-energy surface for large, negative $\ell$.}
  \label{3zeros half charged}
\end{figure}

\begin{figure}[H]
\centering
    \begin{subfigure}[b]{0.4\textwidth}
        \includegraphics[width=0.8\textwidth]{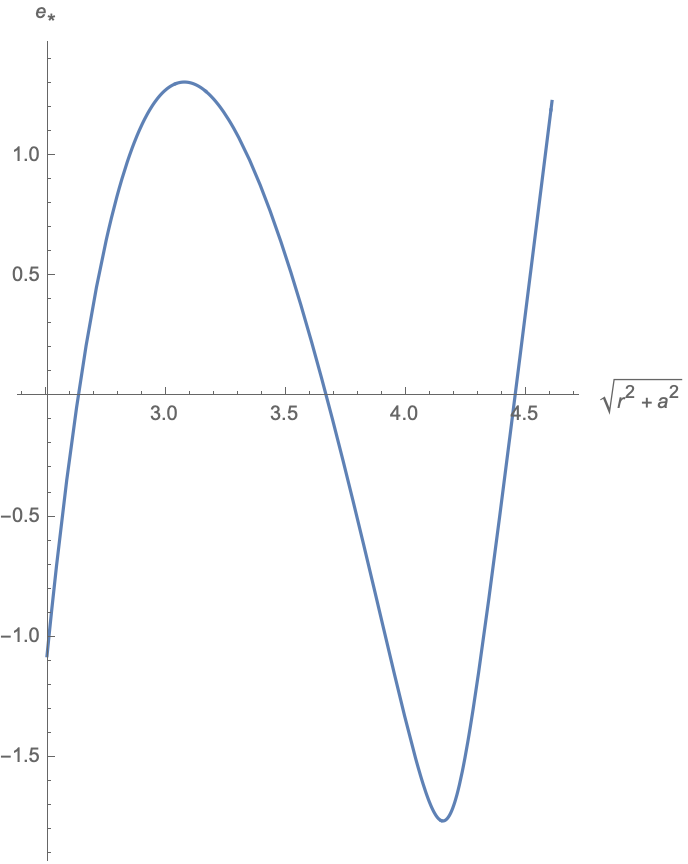}
    \end{subfigure}
    \quad 
    \begin{subfigure}[b]{0.45\textwidth}
        \raisebox{0.07\height}{\includegraphics[width=1.2\textwidth]{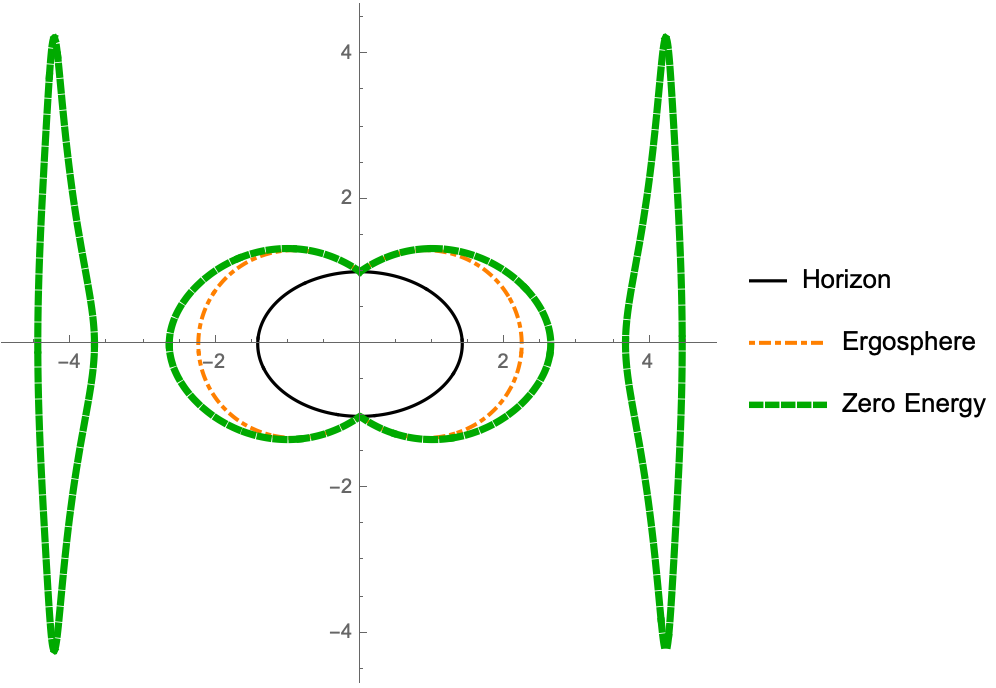}}
    \end{subfigure}   
  \caption{The case with $a=1,\chi_B = -5,\chi_Q =0 ,\ell=-90$. In this case $e_*(r)$ has 3 zeros outside the horizon as shown in the left figure for $\theta=\pi/2$. On the right, a disjoint zero-energy surface at the Wald charge.}
  \label{3zeros Wald}
\end{figure}

\begin{figure}[H]
\centering
    \begin{subfigure}[b]{0.4\textwidth}
        \includegraphics[width=\textwidth, height = \textwidth]{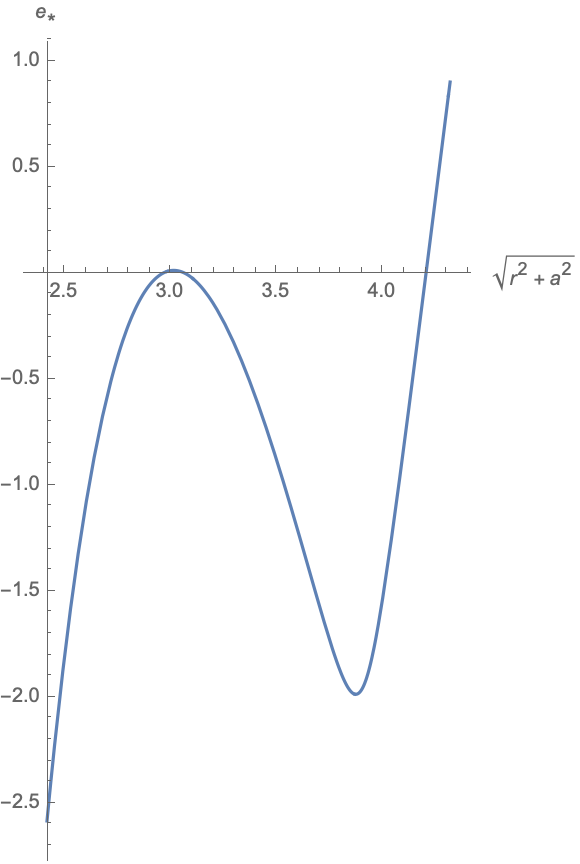}
    \end{subfigure}
    \quad \qquad \qquad
    \begin{subfigure}[b]{0.45\textwidth}
        \raisebox{0.2\height}{\includegraphics[width=\textwidth]{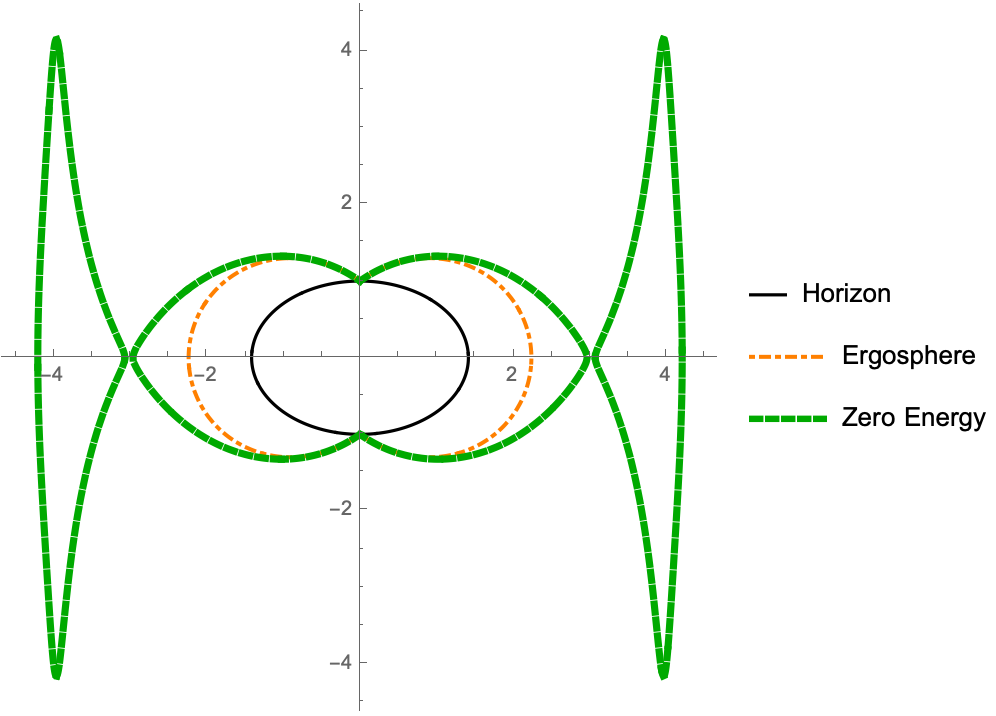}}
    \end{subfigure}   
  \caption{The case with $a=1,\chi_B = -5,\chi_Q =0 ,\ell=-79$. On the left, we show $e_*(r)$ for $\theta=\pi/2$. Compared to Fig. \ref{3zeros Wald}, the two disconnected zero-energy surfaces are closer. If we further decrease $\ell<0$, they will eventually merge as one.}
  \label{3zeros Wald merge}
\end{figure}

An example with $\chi_B>0$ and $\chi_Q<0$ is shown in Fig. \ref{2zeros}.  In this case we have a single zero-energy surface bounding a negative-energy region that is completely detached from the horizon. This allows the possibility that the particle attains a negative energy while getting trapped in this region. A particle with negative energy, perhaps formed by the decay of another particle, will neither fall in or escape. The orbit will be bound in this toroidal region around the black hole.
Numerical investigation shows that in order for $e_*(r)$ to have 3 zeros outside the horizon, $a$ must be close to but not strictly equal to one. An example is shown in Fig. \ref{3zeros}.
As $\chi_Q$ becomes less negative, the allowed range of $\ell>0$ shrinks (and totally disappears as the black hole attains the Wald charge) due to the condition \eqref{b ineq}. The detached negative-energy region will reduce in size and move toward the horizon as shown in Fig. \ref{2zeros charged}. 
Another scenario where we have $\chi_B \chi_Q<0$ is for $\chi_B<0$ and $\chi_Q>0$. Numerical investigations show that when the black hole is totally uncharged ($Q=0$), it is not possible for $e_*(r)$ to have more than 1 zero; in fact, the black hole has to have at least half the Wald charge. Additionally, it is observed that $\chi_B<0$ and $\ell<0$ has to be sufficiently negative. An example is shown in Fig. \ref{3zeros half charged}.
As opposed to the previous case, as $\chi_Q>0$ becomes smaller there are always some values of $\ell<0$ that allow multiple zeros of $e_*(r)$ (for sufficiently large $\chi_B<0$). An example is shown in Fig. \ref{3zeros Wald}. 
As $\ell<0$ gets smaller, the two disconnected zero-energy surfaces will approach each other until they merge eventually. See Fig. \ref{3zeros Wald merge}.

\subsubsection{$\chi_B \chi_Q>0$}

\begin{figure}[H]
\centering
    \begin{subfigure}[b]{0.4\textwidth}
        \includegraphics[width=\textwidth]{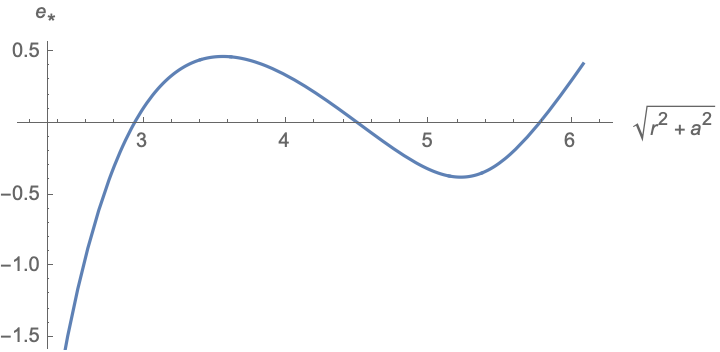}
    \end{subfigure}
    \quad 
    \begin{subfigure}[b]{0.45\textwidth}
        \includegraphics[width=\textwidth]{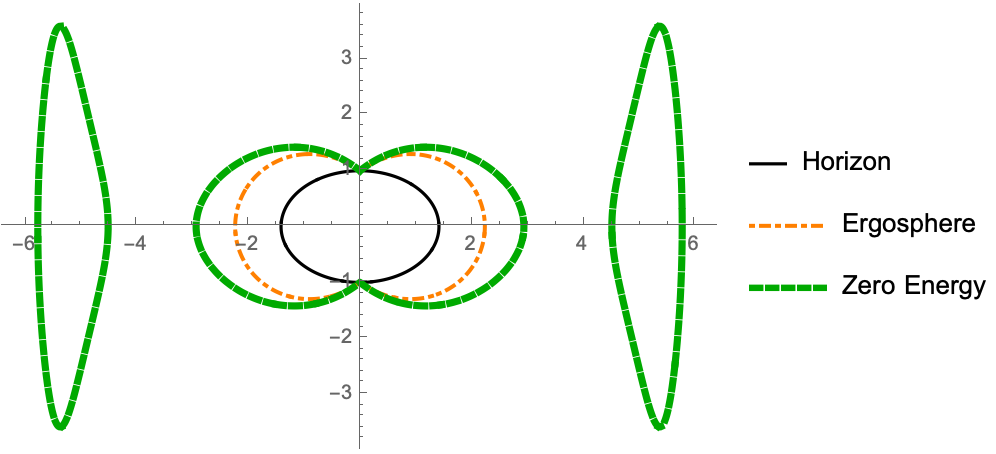}
    \end{subfigure}   
  \caption{The case with $a=1,\chi_B = -1,\chi_Q =-2 ,\ell=-30$. On the left, we show $e_*(r)$ for $\theta=\pi/2$. On the right, the zero-energy surfaces connect.}
  \label{3zeros over}
\end{figure}

Again, this scenario might not be plausible but we consider this for theoretical interest. In this case we have $c_2>c_3$ and the case (i) in \eqref{possibilities} is ruled out. Then we must have $c_3<0$, leading to the condition
\begin{align}\label{c3 ineq}
\chi_B \ell>2\chi_B^2 +\frac{1}{2}>0.
\end{align}
If $\chi_B>0$ and $\chi_Q>0$, this condition says $\ell>0$, but then the condition \eqref{b ineq} cannot be satisfied. We conclude that $e_*(r)$ does not have multiple zeros in this case. Therefore we only need to consider $\chi_B<0$ and $\chi_Q<0$. The inequality \eqref{c3 ineq} becomes
\begin{align}
\ell<2\chi_B +\frac{1}{2\chi_B}<0.
\end{align}
Now, if we want to fall into any of (ii)-(iv) in \eqref{possibilities}, then we cannot have simultaneously $c_1<0$ and $c_2<0$. That is,
\begin{align}
\ell<2\chi_Q+2\chi_B+\frac{1}{2\chi_B}<0\qquad \text{and} \qquad \ell^2-4 \chi_Q^2<0
\end{align}
is not allowed. As it happens, this condition cannot be satisfied for any value of $\ell$. Fig. \ref{3zeros over} shows a case for this scenario.



\section{ORBITS OF PRODUCTS FROM THE PENROSE PROCESS}
\label{Sec:decay}

Let us consider a massive particle of arbitrary charge splitting into two charged massive particles. At the point of split, the 4-momentum and the charge are conserved:
\begin{align}\label{4 mom con}
p_1^\mu=p_2^\mu+p_3^\mu , \qquad q_1 =  q_2 +q_3.
\end{align}
The $t$ and $\phi$ components of the conservation equations are equivalent to
\begin{align}
\mu_1 e_1 =& \mu_2 e_2 +\mu_3 e_3,\label{en con}\\
\mu_1 \ell_1 =& \mu_2 \ell_2 +\mu_3 \ell_3.
\end{align}
Also, each of the particles must obey
\begin{align}\label{mass shell each}
p_i\cdot p_i =-\mu_i^2.
\end{align}
One can consider a more general collisional Penrose process  \cite{1975ApJ...196L.107P,1977ApJ...214..268P,Zaslavskii:2016unn,Berti:2014lva, reviewed in \cite{Schnittman:2018ccg}}, in which the initial state consists of multiple particles. We note however that this is not so different qualitatively and amounts to simply replacing $p_1^\mu$ with the total 4-momentum. 

If we have $e_{*,2}<0$, then we can have super-radiance wherein particle 3 has energy greater than that of the incident particle 1 $\mu_3 e_3=\mu_1 e_1- \mu_2 e_2>\mu_1 e_1$. We write
\begin{align}
u_i^\mu = u_i^t (1,v_i,0,\Omega_i)
\end{align}
where 
\begin{align}
v_i =\frac{dr_i}{dt_i},\quad \Omega_i =\frac{d\phi_i}{dt_i}.
\end{align}
From the definition \eqref{conserved con} of the energy we have
\begin{align}
u_i^t = -\frac{e_i +\frac{q_i}{\mu_i}A_t}{g_{tt}+g_{t\phi}\Omega_i}.
\end{align}
Then, using \eqref{4 mom con}, it is easy to show that
\begin{align}\label{fall in en}
e_3 =\zeta \frac{\mu_1}{\mu_3} \left(e_1 +\frac{q_1}{\mu_1}A_t\right) -\frac{q_3}{\mu_3}A_t,\qquad \zeta =\left( \frac{g_{tt}+g_{t\phi}\Omega_3}{g_{tt}+g_{t\phi}\Omega_1}\right)\left(\frac{\Omega_1-\Omega_2}{\Omega_3-\Omega_2}\right).
\end{align}
Note that this expression is general and only the conservation of momentum is used. In particular, the motion is not assumed to be restricted to the equatorial plane.

The efficiency for energy extraction is
\begin{align}
\epsilon = \frac{\mu_3 e_3-\mu_1 e_1}{\mu_1 e_1}=-\frac{\mu_2 e_2}{\mu_1 e_1}.
\end{align}
This is positive when there is energy extraction. Plugging in \eqref{fall in en}, we have
\begin{align}\label{efficiency}
\epsilon = \zeta-1 + \frac{\zeta q_1-q_3}{\mu_1 e_1}A_t= \zeta-1 + \frac{q_3-\zeta q_1}{\mu_1 u_{1,t} +q_1 A_t }A_t
\end{align}
where all quantities are evaluated at the point of split. The first term $\zeta-1$ corresponds to the mechanical part of the process and the second term proportional to the vector potential $A_t$, corresponds to the electromagnetic extraction. When $A_t$ is small, it falls into the ``low regime'' described in \cite{Tursunov:2019oiq}, where the Penrose process essentially reduces to the mechanical one. In that case the efficiency is simply $\epsilon \approx  \zeta-1$ with maximum value 20.7\%. Otherwise, energy extraction is greatly enhanced or suppressed electromagnetically.\footnote{``Moderate'' and ``Ultra-high-efficient'' regimes in the terminology of \cite{Tursunov:2019oiq}. The latter happens when the parent is neutral, i.e. $q_1 = 0$.} Recalling the definition \eqref{Eq:Ag}, the magnitude of $A_t$ depends strongly on the black hole charge $Q$ and the location where the splitting happens. On the symmetry axis, we have
\begin{align}
A_t (\theta=0)=  -\left (Q-2aMB  \right ) \frac{r}{r^2+a^2},
\end{align}
while on the equatorial plane
\begin{align}
A_t \left (\theta=\frac{\pi}{2} \right )=  -\frac{Q-aMB}{r}.
\end{align}
Therefore, we expect the vector potential term to be important everywhere when the black hole is uncharged and become less important as the black hole charges up. At half the Wald charge ($Q=aMB$), $A_t$ vanishes in the equatorial plane but is still positive along the pole. Past this value, the vector potential will more and more negative near the equatorial plane, while it is still positive along the pole until it vanishes at the Wald charge $Q_W=2aMB$.


Let us study more closely the allowed initial conditions at the point of split. It will be helpful to introduce the $3$-vector notation
\begin{align}
\mathbf{p}^i=(p^r, p^\theta,p^\phi),
\end{align}
so that the spatial components of \eqref{4 mom con} are compactly expressed as
\begin{align}\label{space mom cpt}
\mathbf{p}_1^i=\mathbf{p}_2^i+\mathbf{p}_3^i \ \ .
\end{align}
Suppose we are given the 4-momentum $p_1^\mu$ of the parent particle. Because of the conservation laws, we are not free to choose all components of the daughters' momenta. It is clear that given the spatial momenta $\mathbf{p}_1^i$ of the parent and one of the daughters' $\mathbf{p}_2^i$, the other daughter's spatial momentum $\mathbf{p}_3^i$ is fixed, knocking the initial data that can be chosen for the daughters down to 4. However, the $t$-component $p_2^t$ is determined by the mass shell conditions\ \eqref{mass shell each}. The $t$-component of \eqref{4 mom con} puts two extra constraints on $p_2^t$ and $p_3^t$, and therefore $\mathbf{p}_2^i$ are not all independent. Only two of them can be chosen freely. In fact, it is not difficult to write down the general constraint that must be satisfied by $\mathbf{p}_2^i$. Given charge and angular momentum conservation, the energy conservation equation \eqref{en con} becomes
\begin{align}
\sqrt{\mathbf{p}_1\cdot \mathbf{p}_1+\mu_1^2}=\sqrt{\mathbf{p}_2\cdot \mathbf{p}_2+\mu_2^2}+\sqrt{\mathbf{p}_3\cdot \mathbf{p}_3+\mu_3^2}.
\end{align}
Using \eqref{space mom cpt}, the general constraint that must be satisfied by $\mathbf{p}_2$ is:
\begin{align}\label{General kin constraint}
\left( \mathbf{p}_1\cdot \mathbf{p}_2+\frac{\mu_1^2+\mu_2^2-\mu_3^2}{ 2}\right)^2=\left( \mathbf{p}_1\cdot \mathbf{p}_1+\mu_1^2 \right)\left( \mathbf{p}_2\cdot \mathbf{p}_2+\mu_2^2 \right).
\end{align}
In this notation the energies from \eqref{energy} are simply
\begin{align}
e=& -\bar{q} A_t +\Omega_Z p_\phi+\sqrt{\frac{ \Delta \sin^2\theta}{g_{\phi\phi}}\left( \frac{\mathbf{p}\cdot \mathbf{p}}{\mu^2}+1\right) }\nn\\
=&\chi_Q \left(1-\frac{\Delta \sin^2\theta}{g_{\phi\phi}} \right)-\frac{g_{t\phi}}{g_{\phi\phi}}\ell+\sqrt{\frac{ \Delta \sin^2\theta}{g_{\phi\phi}}\left( \frac{\mathbf{p}\cdot \mathbf{p}}{\mu^2}+1\right) }
\end{align}
giving
\begin{align}\label{general decay en}
e_2=& \chi_{2,Q} \left(1-\frac{\Delta \sin^2\theta}{g_{\phi\phi}} \right)-\frac{g_{t\phi}}{g_{\phi\phi}}\ell_2+\frac{1}{\mu_2}\left( \mathbf{p}_1\cdot \mathbf{p}_2+\frac{\mu_1^2+\mu_2^2-\mu_3^2}{ 2}\right)\sqrt{\frac{ \Delta \sin^2\theta}{g_{\phi\phi}}\frac{1}{\mathbf{p}_1\cdot \mathbf{p}_1+\mu_1^2}}\nn\\
e_3=& \chi_{3,Q} \left(1-\frac{\Delta \sin^2\theta}{g_{\phi\phi}} \right)-\frac{g_{t\phi}}{g_{\phi\phi}}\ell_3+\frac{1}{\mu_3}\left( \mathbf{p}_1\cdot \mathbf{p}_3+\frac{\mu_1^2+\mu_3^2-\mu_2^2}{ 2}\right)\sqrt{\frac{ \Delta \sin^2\theta}{g_{\phi\phi}}\frac{1}{\mathbf{p}_1\cdot \mathbf{p}_1+\mu_1^2}}
\end{align}
where we have used the fact that
\begin{align}
\frac{\mu_1^2+\mu_2^2-\mu_3^2}{2}+\mathbf{p}_1\cdot \mathbf{p}_2=-p_{1,t} p_2^t>0, \quad \frac{\mu_1^2+\mu_3^2-\mu_2^2}{2}+\mathbf{p}_1\cdot \mathbf{p}_3=-p_{1,t} p_3^t>0.
\end{align}
Even though these equations are written in term of $\mathbf{p}_3$, this can be eliminated using \eqref{space mom cpt}. Therefore, these expressions only depend on the location of the point of split and the dot product $\mathbf{p}_1\cdot \mathbf{p}_2$.  

\subsection{Case study: Decay of uncharged parent into particle/antiparticle pair}

As a demonstration, consider an uncharged parent with $p_r=p_\theta=0$ but $\ell\ne 0$ that decays into a particle and antiparticle: $\mu_1=2\mu_2=2\mu_3$ and $\bar q_2=-\bar q_3=\bar q$.
For the preferred range $\chi_B>0,\chi_Q<0$, the positively charged particle can have negative $e$, according to the chart in section \ref{Sec:EMCouplings}, and the negatively charged particles (which has opposite signs for $\chi_B,\chi_Q$) gets the kick in energy. We can choose any values of $\sqrt{r^2+a^2},\theta$ in the negative-energy regions, which sets the orbits of the daughter particles.

\begin{figure}[H]
\centering
     \begin{subfigure}[b]{0.45\textwidth}
        \includegraphics[width=\textwidth]{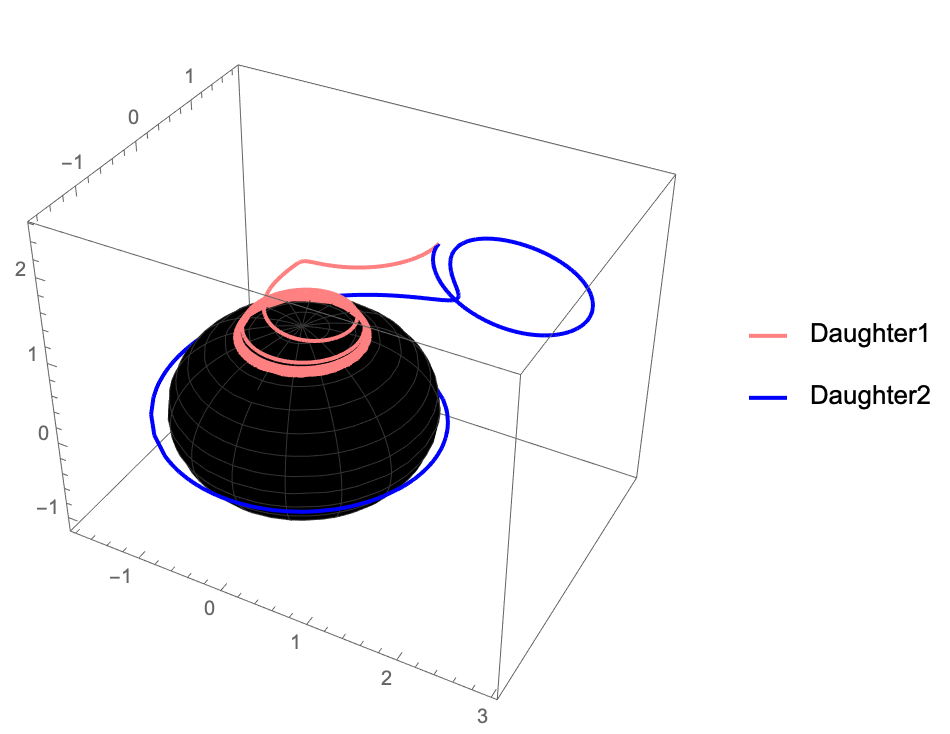}
    \end{subfigure}   
  \caption{The uncharged parent decays at $r=2.8,\theta=\pi/6$ into a negative-energy daughter (Daughter 1) that is trapped within the zero-energy surface in the upper right of Fig. \ref{ZEScompare}. Both the negative-energy daughter (Daughter 1) and the positive-energy daughter (Daughter 2) fall into the black hole.}
  \label{decay orbit 3.2}
\end{figure}

As a simplest example, consider a negative-energy orbit from within the upper-right surface around an uncharged black hole in Fig.\ \ref{ZEScompare} ($\chi_B=10, \chi_Q=-2\chi_B,a=1,\ell=-1$). We choose $r=2.8,\theta=\pi/6$. As shown in Fig.\ \ref{decay orbit 3.2}, the negative-energy particle (red) and the positive-energy particle (blue) both fall into the black hole.

\begin{figure}[H]
\centering
     \begin{subfigure}[b]{0.45\textwidth}
        \includegraphics[width=\textwidth]{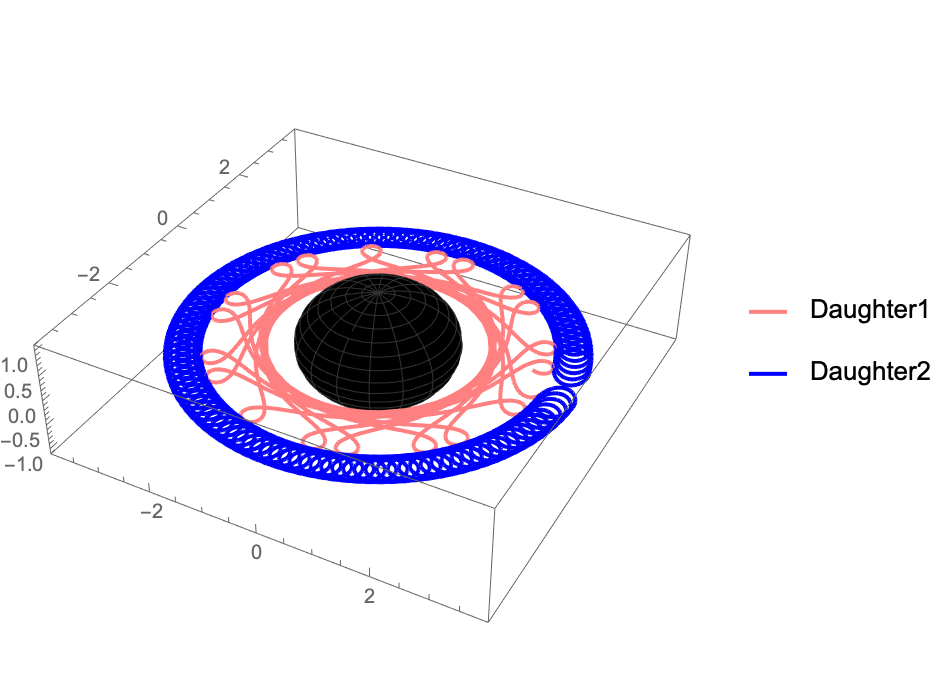}
    \end{subfigure}   
  \caption{The uncharged parent decays into a positively charged, negative-energy daughter (Daughter 1) that is trapped within the zero-energy surface of Fig. \ref{2zeros} and lies in the equatorial plane. The negatively charged, positive-energy daughter (Daughter 2) orbits the black hole.}
  \label{decay orbit 3.5 equatorial}
\end{figure}

As another example, consider orbits in the equatorial plane
corresponding to Fig.\ \ref{2zeros} ($a=1,\chi_B=1, \chi_Q=-2,\ell=5$), we choose $r=2.8,\theta=\pi/2$ for the positively charged particle. As shown in Fig.\ \ref{decay orbit 3.5 equatorial}, the negative-energy particle (red) is trapped in the toroidal region and the positive-energy particle (blue) orbits the black hole.

For a nonequatorial orbit corresponding to Fig.\ \ref{3zeros} ($a=0.97,\chi_B=1, \chi_Q=-2,\ell=5$), we choose the initial values to be $r=3,\theta={\pi}/{5}$.
The negatively charged daughter in blue escapes. The positively charged daughter has negative energy and is confined to the toroidal region of Fig.\ \ref{3zeros}. 

\begin{figure}[H]
\centering
    \begin{subfigure}[b]{0.45\textwidth}
        \includegraphics[width=\textwidth]{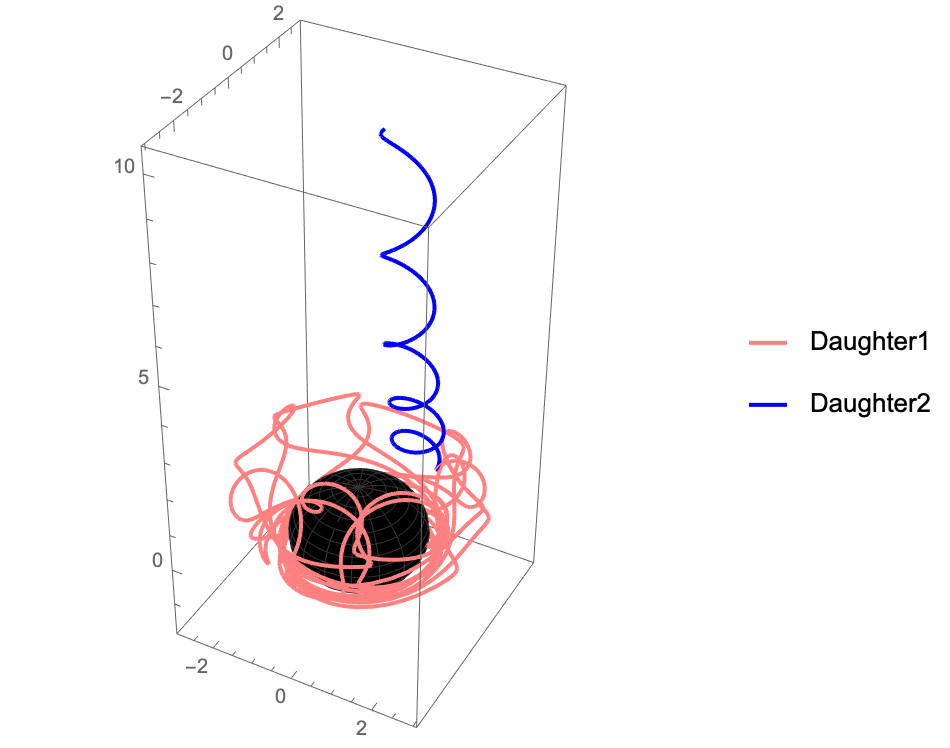}
    \end{subfigure}   
  \caption{The uncharged parent decays into a positively charged, negative-energy daughter (Daughter 1) that is trapped within the zero-energy surface of Fig. \ref{3zeros}. The negatively charged, positive-energy daughter (Daughter 2) escapes.}
  \label{decay orbit 3.6}
\end{figure}

For a nonequatorial orbit corresponding to Fig.\ \ref{3zeros Wald} ($a=1,\chi_B=-5, \chi_Q=0,\ell=-90$), we choose the initial values to be $r=5.9,\theta={\pi}/{4}$.
The positive-energy daughter in blue escapes. The negative-energy daughter is confined to the toroidal region of Fig.\ \ref{3zeros Wald}. 

\begin{figure}[H]
\centering
    \begin{subfigure}[b]{0.45\textwidth}
        \includegraphics[width=\textwidth]{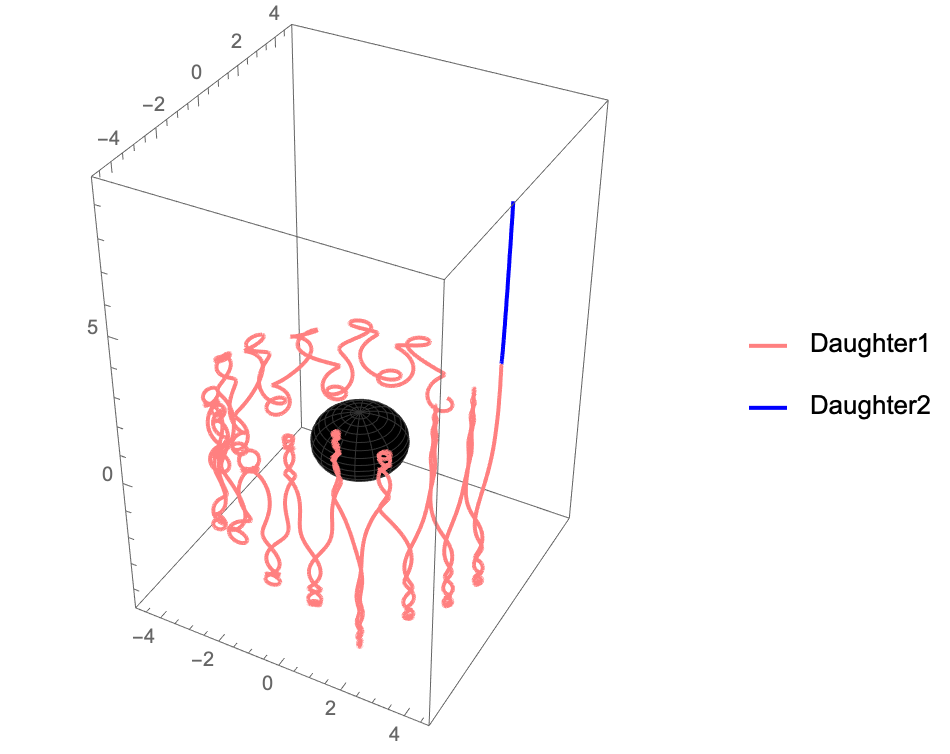}
    \end{subfigure}   
  \caption{The uncharged parent decays into a positively charged, negative-energy daughter (Daughter 1) that is trapped within the zero-energy surface of Fig. \ref{3zeros Wald}. The negatively charged, positive-energy daughter (Daughter 2) escapes.}
  \label{decay orbit 3.9}
\end{figure}

The superradiant particles in Fig. \ref{decay orbit 3.6} and Fig. \ref{decay orbit 3.9} escape along field lines that extend to infinity and so could be particles contributing to the jets. Escaping particles are easily generated. If the split of the parent occurs out of the equator, the positive-energy particle tends to escape along field lines while the negative-energy particle either falls in or is forever within a toroidal zero-energy surface.

\subsubsection{Uncharged stationary parent}

Suppose the parent particle is stationary, $\mathbf{p}_1=0$ so that
\begin{align}
p_{1,t}=-\mu_1e_1-q_1A_t=-\mu_1\bigg(\frac{ \Delta }{g_{\phi\phi}}\bigg)^{1/2}\sin\theta
\end{align}
The parent particle is not stationary with respect to the observer at infinity, because $p_1^\phi=\mu_1\dot{\phi_1}=g^{\phi t} p_{1,t}+g^{\phi\phi}p_{1,\phi}=-g^{\phi t}p_{1,t}\neq 0$. Although this is a subset of the previous section, the equations collapse helpfully. In this case, \eqref{General kin constraint} simply reads
\begin{align}
\mathbf{p}_2\cdot \mathbf{p}_2+\mu_2^2 =\frac{\left (\mu_1^2+\mu_2^2-\mu_3^2\right )}{4 \mu_1^2}\ \ .
\end{align}
Using this constraint, the energies $e_{*,2}$ and $e_{*,3}$ simplify to  
\begin{align}\label{stationary parent en}
e_2=&  \chi_{2,Q} \left(1-\frac{\Delta \sin^2\theta}{g_{\phi\phi}} \right)-\frac{g_{t\phi}}{g_{\phi\phi}}\ell_2+\frac{\mu_1^2+\mu_2^2-\mu_3^2}{ 2\mu_1\mu_2}\bigg(\frac{ \Delta \sin^2\theta}{g_{\phi\phi}}\bigg)^{1/2}\nn\\
e_3=& \chi_{3,Q} \left(1-\frac{\Delta \sin^2\theta}{g_{\phi\phi}} \right)-\frac{g_{t\phi}}{g_{\phi\phi}}\ell_3+\frac{\mu_1^2+\mu_3^2-\mu_2^2}{ 2\mu_1\mu_3}\bigg(\frac{ \Delta \sin^2\theta}{g_{\phi\phi}}\bigg)^{1/2}.
\end{align}
Note that the results \eqref{stationary parent en} are independent of the components $\mathbf{p}_2$ and $\mathbf{p}_3$ and depend only on the location of the point of split. One of the daughters will have negative energy if the split occurs within a zero-energy surface for a given $\ell$.

If a neutral parent decays into a particle and antiparticle, then $\mu_1=2\mu_2=2\mu_3=2\mu$, $\bar{q}_2=-\bar{q}_3=\bar{q}$, $\ell_2=-\ell_3=\ell$, and thus
\begin{align}\label{Eq:eplusminus}
e_2=& \chi_{Q} \left(1-\frac{\Delta \sin^2\theta}{g_{\phi\phi}} \right)-\frac{g_{t\phi}}{g_{\phi\phi}}\ell+
\bigg(\frac{ \Delta \sin^2\theta }{g_{\phi\phi}}\bigg)^{1/2}\nn\\
e_3=&-\chi_{Q} \left(1-\frac{\Delta \sin^2\theta}{g_{\phi\phi}} \right)+\frac{g_{t\phi}}{g_{\phi\phi}}\ell+
\bigg(\frac{ \Delta \sin^2\theta }{g_{\phi\phi}}\bigg)^{1/2}
\end{align}
A comparison of \eqref{Eq:eplusminus} and \eqref{Eq:elong} shows that initially $p_r=p_\theta=0$ and $p_\phi=0\propto\ell-\bar qA_\phi$, fixing
$\ell=\bar q A_\phi$, evaluated at the location of the split. The energies then simplify to:
\begin{align}
\label{Eq:epm}
e_2=& \chi_{Q} \left(g_{tt}+1\right )-\frac{\chi_B}{M}g_{t\phi}+\bigg(\frac{ \Delta \sin^2\theta }{g_{\phi\phi}}\bigg)^{1/2}\nn\\
e_3=&- \chi_{Q} \left(g_{tt}+1\right )+\frac{\chi_B}{M}g_{t\phi}+\bigg(\frac{ \Delta \sin^2\theta }{g_{\phi\phi}}\bigg)^{1/2}
\end{align}

\begin{figure}[H]
\centering
    \begin{subfigure}[b]{0.4\textwidth}
        \raisebox{0.2\height}{\includegraphics[width=\textwidth]{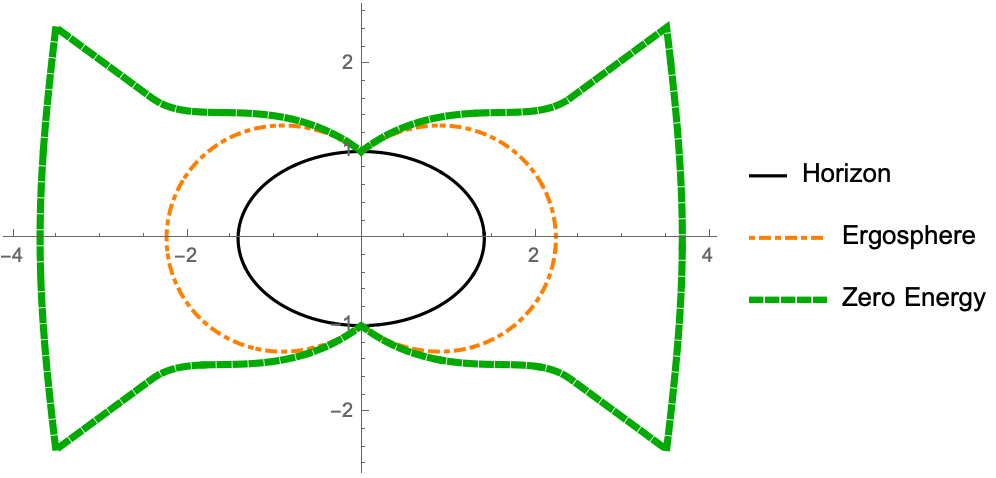}}
    \end{subfigure}
    \qquad
    \begin{subfigure}[b]{0.45\textwidth}
        \includegraphics[width=\textwidth]{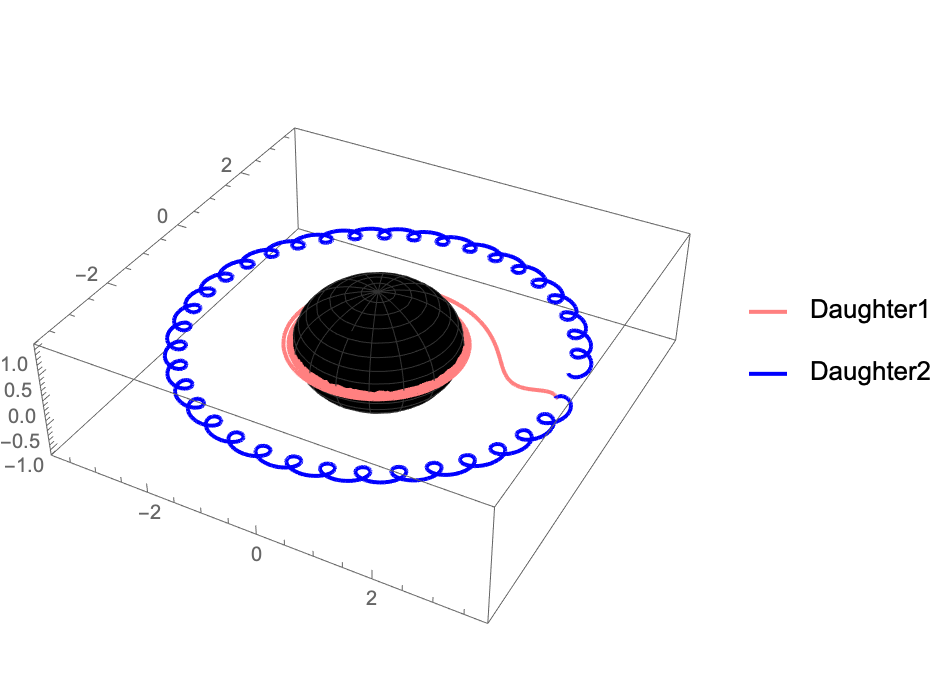}
    \end{subfigure}   
  \caption{The zero-energy surface for $\chi_B=-2,\chi_Q=-1,\ell=-22$. 
  Right: The uncharged parent decays at $r=3,\theta=\pi/2$ into a negative-energy daughter (Daughter 1) and a positive-energy daughter (Daughter 2). Daughter 1 is able to fall into the black hole since it lives within the zero-energy surface shown on the left while Daughter 2 is not able to fall into the black hole.}
  \label{decay orbit 3}
\end{figure}

We  show an equatorial example in Fig. \ref{decay orbit 3}.

\begin{figure}[H]
\centering
    \begin{subfigure}[b]{0.4\textwidth}
        \includegraphics[width=\textwidth,height=1.2\textwidth]{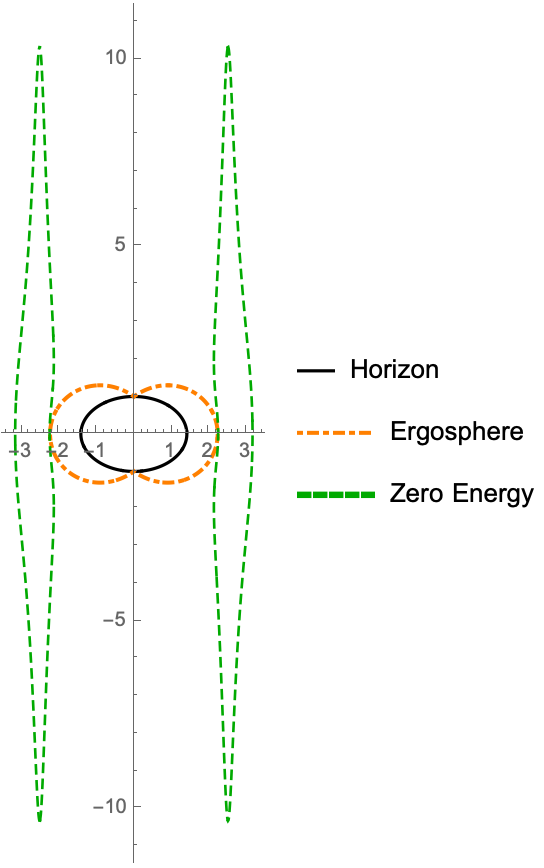}
    \end{subfigure}
    \quad  
    \begin{subfigure}[b]{0.5\textwidth}
        \raisebox{0.1\height}{\includegraphics[width=\textwidth]{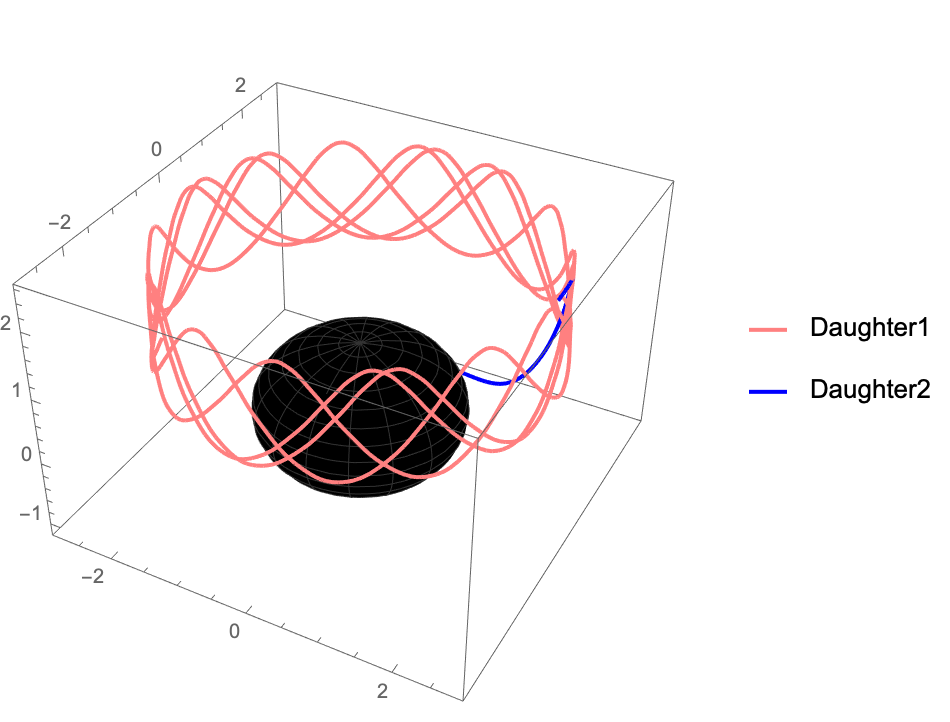}}
    \end{subfigure} 
  \caption{Left: the zero-energy surface for $\chi_B=3,\chi_Q=-5, \ell=18.9$. Right:
the uncharged parent decays at $r=3.5,\theta=\pi/4$ into a negative-energy daughter (Daughter 1) who lives in the zero-energy surface on the left and 
a positive-energy daughter (Daughter 2) who falls in.}
  \label{decay orbit 2}
\end{figure}

Considering a maximally spinning black hole, for the preferred range $\chi_B>0,\chi_Q<0$, it is $e_2$ (for the positive charge) that can be negative, according to the chart in section \ref{Sec:EMCouplings} and $e_{3}$ (for the negative charge) gets the kick in energy, as shown in the example of Fig.\ \ref{decay orbit 2}.

\begin{figure}[H]
\centering
    \begin{subfigure}[b]{0.4\textwidth}
        \includegraphics[width=\textwidth]{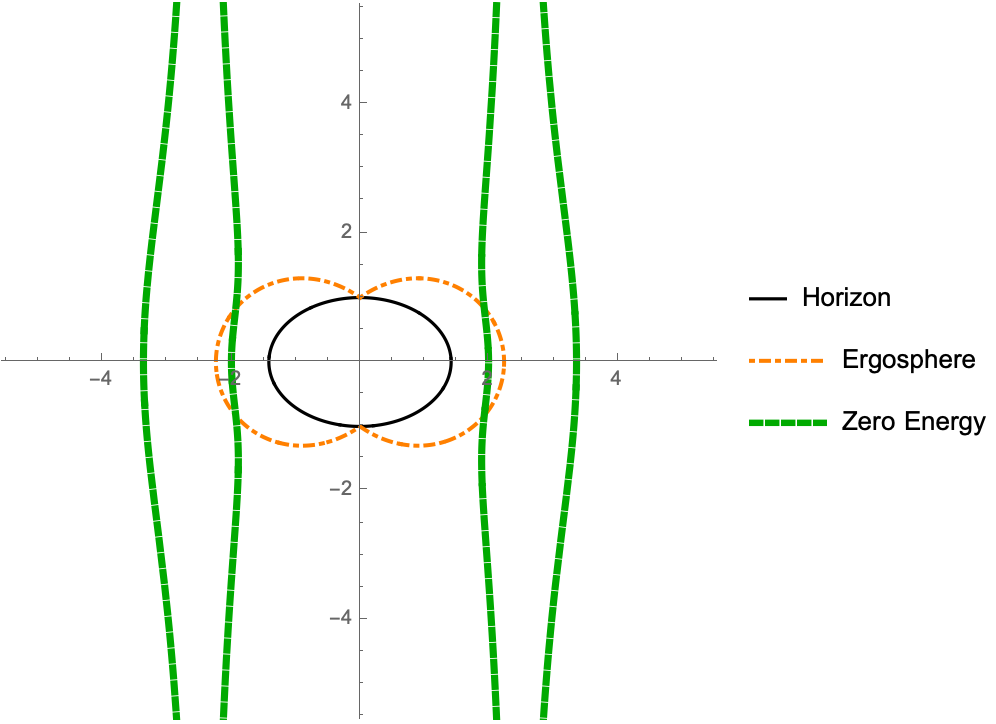}
    \end{subfigure}
    \qquad
    \begin{subfigure}[b]{0.45\textwidth}
        \includegraphics[width=\textwidth]{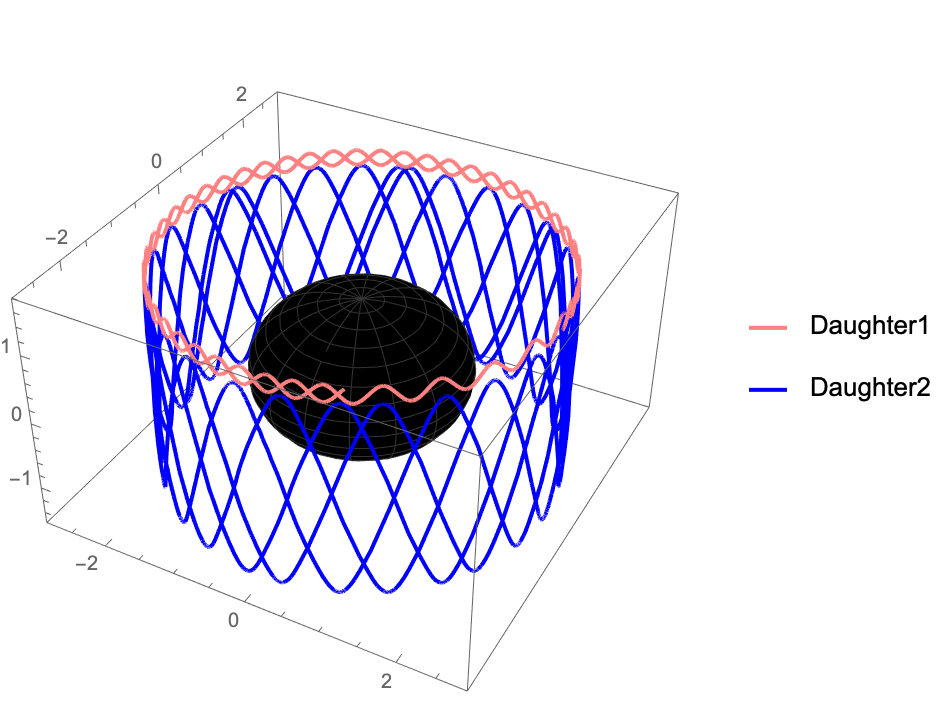}
    \end{subfigure}   
  \caption{Left: the zero-energy surface for $\chi_B=100,\chi_Q=-200,\ell=598.1$. The peak of the surface reaches $z\sim 400$ near the poles.
Right: the uncharged parent decays at $r=2.8,\theta=\pi/3$ into a negative-energy daughter (Daughter 1) and a positive-energy daughter (Daughter 2).}
  \label{decay orbit 4}
\end{figure}

Figure \ref{decay orbit 4} is particularly interesting as it demonstrates the consequence of increasing $\chi_B$ and $\chi_Q$ in magnitude. Both the negative-energy daughter and the positive-energy daughter are trapped in orbit. The energy of the superradiant daughter is 73 times larger than the parent. As the magnitude of $\chi_Q$ approaches values of $10^{10}-10^{21}$, the efficiency will get correspondingly larger.

Our expressions will lend themselves to any of the obvious case studies, such as beta decay, particle-antiparticle collisions, photon emission etc. Any boost in energy a nearby particle experiences can be reflected in the light it emits through any radiative process. The particle itself need not escape to infinity. This raises the interesting prospect that the daughter with negative $e$ may radiate light at enhanced energies too when trapped within a disjoint negative-energy surface, even though the particle itself never escapes to infinity. To address the prospect of light emission would require the full synchrotron radiation-reaction problem as done for instance in \cite{Sokolov:1983gg,1978PhLA...68....1S} or more recently in \cite{Shoom:2015uba,Tursunov:2018erf,Kolos:2020gdc}. We leave this for future work.


\section{IN CLOSING}

We have shown the enhanced power of the electromagnetic Penrose process with regions extended beyond the ergosphere,  including novel toroidal surfaces that trap negative-energy particles in orbit around the black hole. From these regions, tremendous energy can be extracted and delivered to outgoing superradiant particles. 

While we can estimate the efficiency of these process from the effective coupling $\chi_Q$ between the black hole and charged particles, we make no attempt to quantify the probability of energy extraction. Just because a particle {\it can} decay into a trapped negative-energy daughter and a significantly boosted positive-energy radiator, does not mean it {\it will} do so, often or ever. A sophisticated predictive model for the enhanced power of any observable emission -- whether from an accretion disk, a magnetosphere, a black hole battery, or a jet -- would entail detailed numerical modeling as opposed to the clean vacuum solutions exploited here. Perhaps more fruitful would be to scan observations for anomalous augmented power and extrapolate from there.

We are encouraged by this era of precision black hole astrophysics. In the range of stellar to intermediate mass black holes, a network of observatories promises multimessenger counterparts to gravitational-waves. In the supermassive range, the Event Horizon Telescope project captures detailed observations of emission mechanisms in real time. With such meticulous detections emerging,  electromagnetic Penrose processes could leave  observable imprints on black holes and their luminous environments. It would be intriguing to consider, for instance, implications of the generalized Penrose process on polarization of light emitted near the event horizon of M87*\cite{MagFieldEHT}.


\section*{Acknowledgments} 

We thank Roman Berens and Prakruth Adari for collaborative conversations. 
AL was supported in part by the U.S. Department of Energy grant de-sc0011941. JL is supported in part by the Tow Foundation. 


%
\bibliographystyle{unsrt}
\bibliography{main}

\end{document}